\documentclass[12pt, draftclsnofoot, onecolumn]{IEEEtran}

\usepackage{booktabs}
\usepackage[T1]{fontenc}
\usepackage{multirow}
\usepackage{siunitx}

\usepackage{amsmath,amssymb,amsfonts}
\usepackage[final]{graphicx}
\usepackage{xcolor}
\usepackage{flushend}
\usepackage[normalem]{ulem} 

\usepackage[numbers,sort&compress]{natbib}	

\newtheorem{proposition}{Proposition}
\newtheorem{lemma}{Lemma}

\def \M{\mathcal{M}}

\def \K {{\mathcal K }}

\def\M{ {\mathcal{M}} }
\def\K{ {\mathcal{K}} }
\def\S{ {\mathcal{S}} }
\def\W{ {\mathcal{W}} }

\makeatletter
\def\blfootnote{\xdef\@thefnmark{}\@footnotetext}
\makeatother

\begin{document}
\title{\Huge A Tractable Product Channel Model for Line-of-Sight Scenarios}

\author{
{{U. Fernandez-Plazaola}, L. Moreno-Pozas, F. J. Lopez-Martinez},\\ J. F. Paris, E. Martos-Naya and  {J. M. Romero-Jerez}}

\markboth{submitted to IEEE Journal}%
{U. Fernandez-Plazaola \MakeLowercase{\textit{et al.}}: A Tractable Product Channel Model for Line-of-Sight Scenarios}

\vspace{-2mm}
\maketitle
\begin{abstract}
We present a general and tractable fading model for line-of-sight (LOS) scenarios, which is based on the product of two independent and non-identically distributed $\kappa$-$\mu$ shadowed random variables. Simple closed-form expressions for the probability density function, cumulative distribution function and moment-generating function are derived, which are as tractable as the corresponding expressions derived from a product of Nakagami-$m$ random variables. This model simplifies the challenging characterization of LOS product channels, as well as combinations of LOS channels with non-LOS ones. We leverage these results to analyze performance measures of interest in the contexts of wireless powered and backscatter communications, where both forward and reverse links are inherently of LOS nature, as well as in device-to-device communications subject to composite fading. In these contexts, the model shows a higher flexibility when fitting field measurements with respect to conventional approaches based on product distributions with deterministic LOS, together with a more complete physical interpretation of the underlying propagation characteristics.
\end{abstract}

\vspace{0mm}

\blfootnote{\noindent  U. Fernandez-Plazaola, F. J. Lopez-Martinez,  J. F. Paris, and~ E. Martos-Naya are with Dpto. Ingenier\'ia de Comunicaciones, Universidad de Malaga, Malaga 29071, Spain. E-mail: \{{\rm unai,fjlopezm,paris,eduardo}\}@ic.uma.es. L. Moreno-Pozas is with Dept. Electronic \& Computer Engineering at the Hong Kong University of Science and Technology (HKUST), Clear Water Bay, Kowloon, Hong Kong. E-mail: {\rm eelaureano@ust.hk}. J. M. Romero-Jerez is with Dpto. Tecnolog\'ia Electr\'onica, Universidad de Malaga, Malaga 29071, Spain. E-mail: {\rm romero@dte.uma.es}.\\
\indent This work has been funded by the Spanish Government
and the European Fund for Regional Development FEDER TEC2014-57901-R and TEC2017-87913-R. This work was presented in part at the 15th International Symposium on Wireless Communication Systems (ISWCS) 2018. This work has been submitted to the IEEE for possible publication. Copyright may be transferred without notice, after which this version may no longer be accessible.
}

\IEEEpeerreviewmaketitle

\section{Introduction}

The statistical characterization of products of random variables (RVs)
plays an important role in a wide range of applications, including statistical testing \cite{Springer1967}, hydrology \cite{Nadarajah2007}, cosmology \cite{Sumitomo2012} and wireless communications \cite{Salo2006}. Even for the simplest case where only two RVs are considered, i.e. $Z=XY$, with $X$ and $Y$ being independent, the statistical characterization 
of $Z$ is usually much more involved that the individual distributions of $X$ and $Y$.

In wireless communications, {and more specifically in the context of wireless channel modeling, $Z$ is usually} referred to as \textit{product channel} or {\textit{cascaded channel}}. The product channel naturally arises in the context of communication systems assisted by relays \cite{Karagiannidis2007}, when modeling propagation effects such as keyholes \cite{Chizhik2002}, diffraction \cite{Erceg1997} and composite fading \cite{Sofotasios2016}, or turbulence-induced scintillation in free-space optical communications \cite{Andrews2001}. Further scenarios on which the product channel characterization is essential also include wireless powered communications (WPC) \cite{Zhong2015,Van2016,Alcaraz2017} and backscatter communications \cite{Kim2003,Griffin2008}. Thus, the statistical characterization of the product channel is of paramount relevance for understanding the performance limits of wireless communication systems operating in these scenarios, either in line-of-sight (LOS) or non-LOS (NLOS) propagation conditions.

Taking a deeper look into the literature, one alternative in many scenarios {relies on} approximating the resulting product channel as a product of two NLOS fading channels \cite{Kim2003, Griffin2008,Zhong2015,Van2016,Alcaraz2017} regardless of the propagation conditions. The main motivation behind this approach is defining a tractable framework, since the characterization of LOS product (or LOSxLOS) channels is very challenging and cannot be given in closed-form, but in terms of double infinite sums of special functions \cite{Odonoughue2012}. Therefore, for the sake of tractability, some authors have approximated both LOS links as NLOS ones, which have closed-form characterization \cite{Erceg1997, Karagiannidis2007}. Indeed, using the Nakagami-$m$ distribution for approximating the Rician distribution is a classical approach which can simplify the characterization of some scenarios \cite{Nakagami1960,AlouiniBook}. In other scenarios, however, the results derived therein can give inaccurate approximations when the propagation conditions are clearly LOS. Moreover, the approximation of the Rician distribution through a Nakagami-$m$ distribution has severe limitations related to the different diversity order of such distributions \cite{Wang2003}; this is especially relevant in the high signal-to-noise-ratio (SNR) regime, which is of key importance when analyzing system performances. {Recent results extend the characterization of the LOS product channel distributions to the $\kappa$-$\mu$ case \cite{Bhargav2018}, as well as combinations of LOS and NLOS cases involving products of $\kappa$-$\mu$ and $\eta$-$\mu$ or $\alpha$-$\mu$ channels \cite{Silva2018}. In all these works, the probability density function (PDF) and CDF have complicated forms similar to those involving Rician product channels.}

With all these considerations, the literature is {lacking LOS} product channel models which are analytically tractable. Since the complexity of previous results for LOS product channel models are mainly due to the challenge posed by considering a product of two Rician RVs, we here propose to characterize product channels by means of a more general distribution, which in turn will help simplifying the problem. We will introduce a product channel model based on the $\kappa$-$\mu$ shadowed fading distribution \cite{Paris2014,Cotton2015}, built as the product of two independent and non-identically distributed (INID) $\kappa$-$\mu$ shadowed RVs with integer fading parameters. This includes the Rician shadowed product and the Rician product models as a special cases. For the sake of shorthand notation, we will refer to this new fading distribution as the $\mathcal{P}$-distribution (where $\mathcal{P}$ stands for product). The $\mathcal{P}$-distribution inherits the same physical characteristics inherent to the $\kappa$-$\mu$ shadowed fading model from which it originates, and most notably, its ability to model random fluctuations on the LOS components. {Although we here focus on simplifying the analysis of Rician product channels and their generalizations, the results here presented for the $\kappa$-$\mu$ shadowed product channel can unify the analysis of a vast set of product models as special cases, including LOSxLOS product channel based on the $\kappa$-$\mu$ distribution, as well as LOSxNLOS and NLOSxNLOS product channels based on the Nakagami-$m$ and Rayleigh distributions \cite{Laureano2015}}. The usefulness of this new distribution, both in terms of tractability and improved fit to field measurements, is exemplified in the contexts of WPC, backscatter communications and device-to-device (D2D) communications.

The remainder of this paper is structured as follows. The chief probability functions of the $\mathcal{P}$-distribution is introduced in Section \ref{statchar}. {Then, the application of this distribution to several scenarios of interest is addressed in Section \ref{Applications}: WPC, backscatter and D2D communications. Numerical results are presented in Section \ref{NUM}, whereas} main conclusions are drawn in Section \ref{Conclusions}.

\section{Statistical Characterization}
\label{statchar}
In this section, we will derive the chief probability functions characterizing the $\mathcal{P}$-distribution, which is built from the product of two INID $\kappa$-$\mu$ shadowed RVs. Throughout this paper, we will consider the distributions associated to the power envelope in $\kappa$-$\mu$ shadowed fading channels (or equivalently, the instantaneous receive SNR $\gamma$). The distribution of the received signal envelope $r$ can be easily computed through a simple change of variables, assuming that $\gamma\propto r^2$.

\subsection{Preliminary results}
We will first present some preliminary results that will become relevant for the following derivations.

\begin{lemma}[SNR distribution under $\kappa$-$\mu$ shadowed fading with integer parameters \emph{\cite{Lopez2017}}]
\label{statchar_lema1}
Let $\gamma$ be a squared $\kappa$-$\mu$ shadowed random variable with mean $\bar\gamma$ and shape parameters $\kappa$, $\mu$ and $m$ \cite{Paris2014}. If the parameters $\mu$ and $m$ are restricted to be positive integers, then for any arbitrary non-negative real $\kappa$ the PDF and CDF of $\gamma$ are given by \cite[eq.~(4-10)]{Lopez2017}

\begin{equation}
\label{statchar_eq_1}
f_{\S} \left( x \right) = \sum\limits_{j = 0}^M {C_j
\underbrace{\frac{{{x}^{m_j  - 1} } e^{ - \frac{{x}}{{\Omega _j }}}   }
{\Omega _j^{m_j }   {\left( {m_j  - 1} \right)!}}
 }_{f_{\K} ({\Omega _j ;m_j ;x}) }},
\end{equation}
\begin{equation}
\label{statchar_eq_2}
F_{\S} \left( x \right) = 1 - \sum\limits_{j = 0}^M {C_j e^{ - \frac{x}
{{\Omega _j }}} \sum\limits_{r = 0}^{m_j  - 1} {\frac{1}
{{r!}}\left( {\frac{x}
{{\Omega _j }}} \right)^r } },
\end{equation}
where $M$ and the set of parameters $\{C_j,m_j,\Omega_j\}_{j=1,...,M}$ are expressed
in terms of $\bar\gamma$, $\kappa$, $\mu$ and $m$ according to {Table \ref{table01}}. In (\ref{statchar_eq_1}),
$f_{\K}(\cdot)$ represents the PDF of a squared Nakagami-$m$ distribution, (i.e. a Gamma
distribution).
\end{lemma}

\begin{table*}[t]
  \renewcommand{\arraystretch}{3}
\centering
{
\caption{Parameter values for the SNR distribution under $\kappa$-$\mu$ shadowed fading with integer $\mu$ and $m$,}
\label{table01}
\begin{tabular}{|c|c|}
\hline\hline
Case $\mu>m$ & Case $\mu \leq m$ \\ \hline\hline 
$M=\mu$ & $M=m-\mu$  \\ \hline
 $C_i=\begin{cases} 
      0 & i=0 \\
       \left( { - 1} \right)^m \binom{m+i-2}{i-1}\times \left[ {\frac{m}
{{\mu \kappa  + m}}} \right]^{ m} \left[ {\frac{{\mu \kappa }}
{{\mu \kappa  + m}}} \right]^{ - m - i + 1}  & 0<i\leq \mu-m \\
      \left( { - 1} \right)^{i-\mu+m - 1} \binom{i-2}{i-\mu+m-1} \times \left[ {\frac{m}
{{\mu \kappa  + m}}} \right]^{i-\mu+m - 1} \left[ {\frac{{\mu \kappa }}
{{\mu \kappa  + m}}} \right]^{-i + 1}  & \mu-m < i \leq \mu 
   \end{cases}$
 & $C_i=\binom{m-\mu}{i}\left[ {\frac{m}
{{\mu \kappa  + m}}} \right]^i \left[ {\frac{{\mu \kappa }}
{{\mu \kappa  + m}}} \right]^{m - \mu  - i}$  \\ \hline
\ $m_i=\begin{cases} \mu-m-i+1, & 0\leq i\leq \mu-m \\
   \mu-i+1 & \mu-m < i \leq \mu 
   \end{cases}$ & $m_i=m-i$  \\ \hline
    $\Omega_i=\begin{cases}  \frac{{\bar \gamma }}
{{\mu \left( {1 + \kappa } \right)}}, & 0\leq i\leq \mu-m \\
   \frac{{\mu \kappa  + m}}
{m}\frac{{\bar \gamma }}
{{\mu \left( {1 + \kappa } \right)}} & \mu-m < i \leq \mu 
   \end{cases}$ & $\Omega_i=\frac{{\mu \kappa  + m}}
{m}\frac{{\bar \gamma }}
{{\mu \left( {1 + \kappa } \right)}}$  \\ \hline
\end{tabular}
}
\end{table*}

\vspace{2mm}
{Note here} that, according to Lemma \ref{statchar_lema1}, the distribution of the SNR under \mbox{$\kappa$-$\mu$} shadowed fading with integer parameters $m$ and $\mu$ can be expressed as a finite mixture of squared Nakagami-$m$ (or Gamma) distribution. To theoretically obtain the Rician distribution as special case, we need to set $\mu=1$ and let $m\rightarrow\infty$. 
\vspace{2mm}

\begin{lemma}[Product of Two Squared Nakagami-$m$ RVs]
\label{statchar_lema2}
Let $Z_{\rm Nak}=W\hat{W}$ be the product of two INID squared Nakagami-$m$ random variables $W$ and $\hat{W}$ with means $\Omega$ and $\hat{\Omega}$, where the corresponding shape parameters $m$ and $\hat{m}$ are arbitrary positive integer numbers. Then, the corresponding PDF and CDF
are given by
\begin{equation}
\label{statchar_eq_3}
f_{\Gamma\Gamma}  \left( x \right) = \frac{2 \; x^{\frac{{m + \hat{m} }}{2} - 1} }
{{\Gamma \left( {m } \right)\Gamma \left( \hat{m } \right)}{\left(\Omega \hat{\Omega} \right)^{\frac{{m  + \hat{m} }}{2}}}}
 K_{m  - \hat{m} } \left( {\sqrt {\frac{{4 x}}
{{\Omega \hat{\Omega} }}} } \right),
\end{equation}
\begin{equation}
\label{PhuTuan}
\begin{gathered}
F_{\Gamma\Gamma}(x) = 1 - \sum\limits_{k = 0}^{m  - 1} {\frac{2}{{k!\Gamma \left( \hat{m } \right)}}} \left( {\frac{x}{\Omega \hat{\Omega} }} \right)^{\frac{{k + \hat{m} }}{2}} K_{\hat{m}  - k} \left( {\sqrt {\frac{4 \; x}{{\Omega \hat{\Omega} }}} } \right)
\end{gathered}
\end{equation}
where $K_{\nu}$ is the modified Bessel function of the second kind, and $\Gamma(\cdot)$ is the Gamma function.
\end{lemma}
\begin{IEEEproof}
The PDF follows from the corresponding expression given in \cite{Karagiannidis2007}{, or as a special case of \cite[eq. (16)]{Leonardo2015},} after performing a simple random variable transformation of the type $Y=X^2$. The CDF is a particular case of \cite[eq. 8]{Van2016}.
\end{IEEEproof}

The distribution described in Lemma \ref{statchar_lema2} is essentially a Gamma-Gamma ($\Gamma\Gamma$) distribution, up to a trivial re-scaling by $\Omega$. 
For the sake of notation simplicity, in this work we will refer to this distribution as a $\Gamma\Gamma$ distribution. 
\subsection{Main results}

By means of the previous results and considerations, we will now characterize the distribution of the product of two INID $\kappa$-$\mu$ shadowed fading variables with integer fading parameters.

\begin{proposition}[The $\mathcal{P}$-distribution as a finite mixture of $\Gamma\Gamma$ distributions]
\label{statchar_theo1}
Let $Z$ be the product of two INID squared \mbox{$\kappa$-$\mu$} shadowed random variables $X$ and $\hat X$ with means $\bar\gamma$ and $\widetilde{\gamma}$. The
corresponding shape parameters $\kappa$ and $\hat\kappa$ are arbitrary non-negative real numbers and the remainder shape parameters $\mu$ and $m$ for $X$, and $\hat\mu$ and $\hat m$ for $\hat X$ are positive integers.
Under these conditions, $Z=X\hat X$ is distributed as a $\Gamma\Gamma$ finite mixture with the following PDF
\begin{equation}
\label{statchar_eq_5}
\begin{gathered}
  f_Z \left( z \right) = \sum\limits_{j = 0}^M {\sum\limits_{h = 0}^{\hat M} {C_j \hat C_h }  \times }  \underbrace{\frac{2 \; z^{\frac{{m_j  + \hat m_h }}{2} - 1} }
{{\Gamma ( {m_j } )\Gamma \left( {\hat m_h } \right)}{\left(\Omega _j \Omega _h \right)^{\frac{{m_j  + \hat m_h }}{2}}}}
 K_{m_j  - \hat m_h } \left( {\sqrt {\frac{{4 z}}
{{\Omega _j \hat \Omega _h }}} } \right)}_{
f_{\Gamma \Gamma } \left( {z;\left\{ {\Omega _j ,m_j } \right\};\left\{ {\hat \Omega _h ,\hat m_h } \right\}} \right)
} \hfill \\
\end{gathered}
\end{equation}
where the parameters $M$ and $\{C_j,m_j,\Omega_j\}_{j=1,...,M}$ { are computed thanks to Table \ref{table01}. For the
parameters ${\hat M}$ and $\{{\hat C}_h,{\hat m}_h,{\hat\Omega}_h\}_{h=1,...,\hat M}$, we can also employ Table \ref{table01} when one substitutes $\bar{\gamma}$, $\kappa$, $\mu$ and $m$ by 
$\widetilde{\gamma}$, $\hat\kappa$, $\hat\mu$ and $\hat m$, respectively.}
\end{proposition}
\begin{IEEEproof}
See Appendix \ref{appendix:pdf}.
\end{IEEEproof}

Proposition \ref{statchar_theo1} states that the $\mathcal{P}$-distribution can be expressed in closed-form as a finite sum of well-known special functions. For $\mu=\hat\mu=1$, a simplified version of the $\mathcal{P}$-distribution is obtained as a by-product, built as the product of two INID Rician shadowed RVs. In any case, the $\mathcal{P}$-distribution is more general \textit{and} simpler than the Rician product distribution (which is but a special case for ${\mu=\hat\mu=1}$ and sufficiently large $m$ and $\hat m$). Hence, we advocate for its use as the reference product channel model in a communication-theoretic context. This will be later supported by both theoretical and practical evidences in different scenarios of interest.

{The effect of the fading parameters in this model is inherited from the underlying $\kappa$-$\mu$ shadowed distributions from which its built, and has been well-documented in the related literature \cite{Paris2014,Cotton2015,Laureano2015,Lopez2017}. In the next set of figures, we exemplify the impact of the different parameters of the $\mathcal{P}$-distribution on the shape of the PDF.}

\begin{figure}[t]
\centering
  \includegraphics[width=0.61\columnwidth]{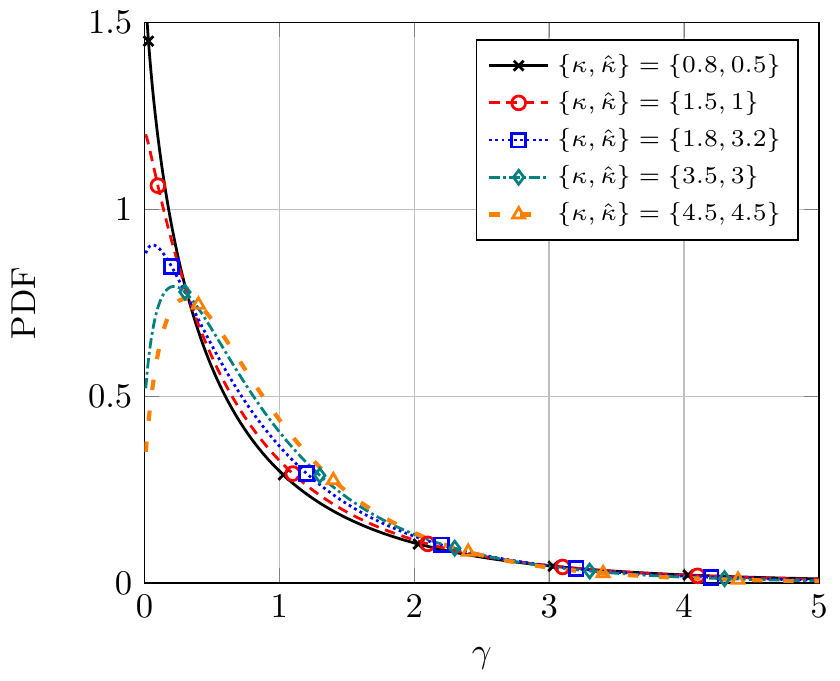}
          \caption{{Normalized power envelope $\mathcal{P}$-distribution for different values of $\kappa$ and $\hat{\kappa}$, with $\mu=1$, $\hat{\mu}=2$, $m=5$ and $\hat{m}=10$. Solid lines correspond to the exact PDF derived from eq. (5) in the paper, markers correspond to Monte Carlo simulations.}   } 
	\label{Figkappas}
\end{figure}

\begin{figure}[t]
\centering
  \includegraphics[width=0.61\columnwidth]{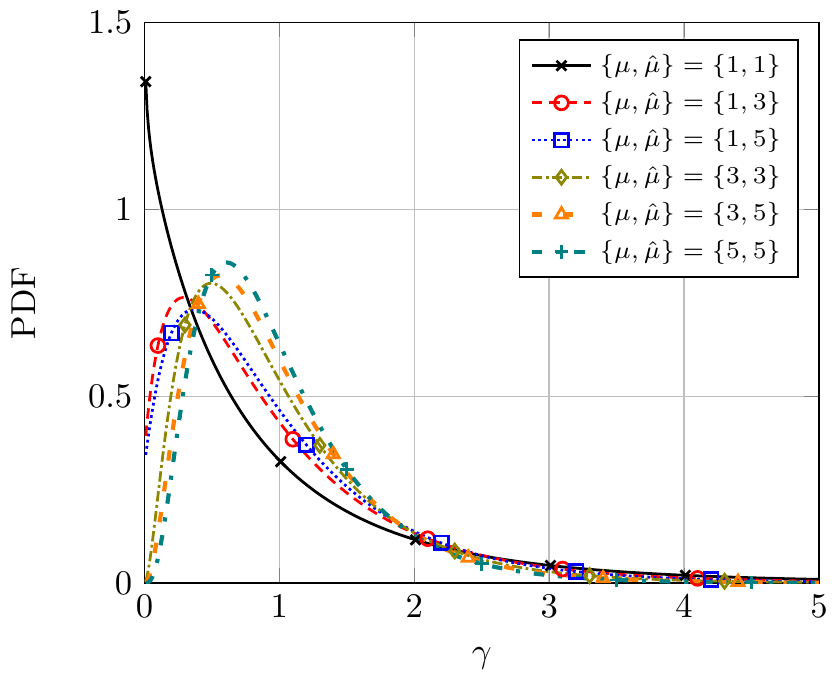}
          \caption{{Normalized power envelope $\mathcal{P}$-distribution for different values of $\mu$ and $\hat{\mu}$, with $\kappa=4$, $\hat{\kappa}=2$, $m=5$, and $\hat{m}=10$. Solid lines correspond to the exact PDF derived from eq. (5) in the paper, markers correspond to Monte Carlo simulations.}}  	\label{Figmus}
\end{figure}

\begin{figure}[t]
\centering
  \includegraphics[width=0.61\columnwidth]{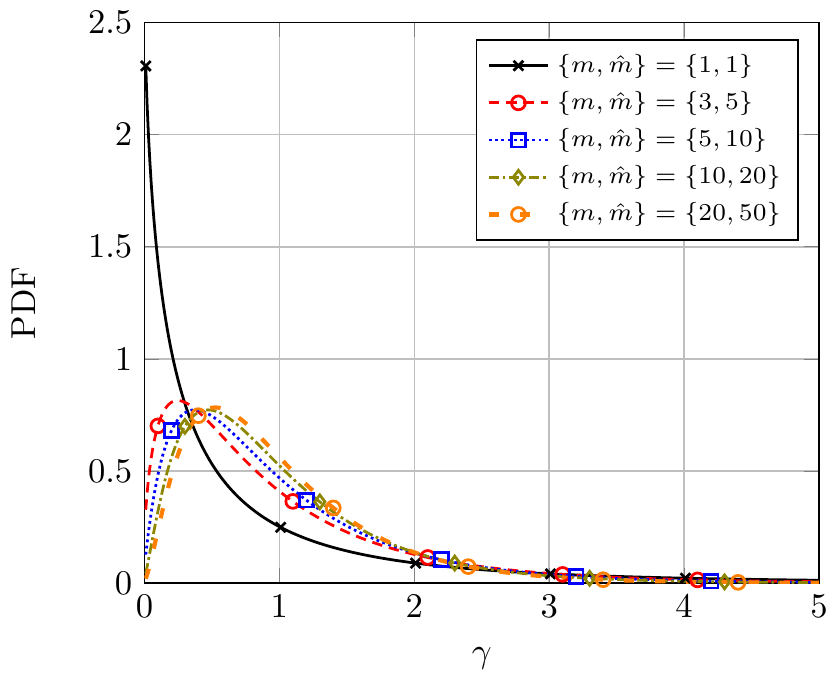}
          \caption{{Normalized power envelope $\mathcal{P}$-distribution for different values of $m$ and $\hat{m}$, with $\mu=1$, $\hat{\mu}=2$, $\kappa=10$ and $\hat{\kappa}=3$. Solid lines correspond to the exact PDF derived from eq. (5) in the paper, markers correspond to Monte Carlo simulations.}}
	\label{Figms}
\end{figure}
{
We first set fixed values for the parameters $\mu=1$, $\hat{\mu}=2$, $m=5$, $\hat{m}=10$ and let $\kappa$ and $\hat{\kappa}$ vary to generate Fig. \ref{Figkappas}. In this situation, as the magnitude of the specular components associated to LOS propagation becomes dominant (i.e. a larger value of $\kappa$ or $\hat{\kappa}$), the probability of having a low SNR is decreased and the shape of the PDF around zero changes to reflect such behavior.

In Fig. \ref{Figmus}, the parameter values $\kappa=4$, $\hat{\kappa}=2$, $m=5$, and $\hat{m}=10$ are set and now $\mu$ and $\hat{\mu}$ change. As the number of clusters rises, which implies a milder multipath fading, the probability of having a better SNR is increased. Again, the shape of the PDF is modified accordingly.

Finally, we consider $\mu=1$, $\hat{\mu}=2$, $\kappa=10$, $\hat{\kappa}=3$, and modify $m$ and $\hat{m}$. When the dominant specular components severely fluctuate (i.e. very low values of $m$ and $\hat{m}$), the probability of having a low SNR grows. In this situation, the shape of the PDF resembles that of the case with low $\{\kappa,\hat{\kappa}\}$ in Fig. \ref{Figkappas} or low $\{\mu,\hat{\mu}\}$ in Fig. \ref{Figmus}. As these fluctuations are reduced, i.e. $m$ and $\hat{m}$ are increased, the probability of having a better SNR is also increased.
}
\begin{proposition}[CDF of the $\mathcal{P}$-distribution as a finite mixture]
\label{statchar_col1}
Let $Z$ be the product of two INID squared $\kappa$-$\mu$ shadowed random variables $X$ and $\hat X$ with means $\bar\gamma$ and $\widetilde{\gamma}$. The corresponding shape parameters $\kappa$ and $\hat\kappa$ are arbitrary non-negative real numbers and the remainder shape parameters $\mu$ and $m$ for $X$, and $\hat\mu$ and $\hat m$ for $\hat X$ are positive integers. Under these conditions $Z=X\hat X$ has the following CDF

\begin{equation}
\label{statchar_eq_9}
\begin{gathered}
  F_Z \left( z \right) = \sum\limits_{j = 0}^M {\sum\limits_{h = 0}^{\hat M} {C_j \hat C_h }  \times }  \underbrace{\left[ {1 - \sum\limits_{k = 0}^{m_j  - 1} {\frac{2}{{k!\Gamma \left( {\hat m_h } \right)}}} \left( {\frac{z}{{\Omega _j \hat \Omega _h }}} \right)^{\frac{{k + \hat m_h }}{2}}  \times K_{\hat m_h  - k} \left( {\sqrt {\frac{{4\:z}}{{\Omega _j \hat \Omega _h }}} } \right)} \right]
}_{
F_{\Gamma \Gamma } \left( {z;\left\{ {\Omega _j ,m_j } \right\};\left\{ {\hat \Omega _h ,\hat m_h } \right\}} \right)
} \hfill 
\end{gathered}
\end{equation}
where the parameters $M$ and $\{C_j,m_j,\Omega_j\}_{j=1,...,M}$, ${\hat M}$ and $\{{\hat C}_h,{\hat m}_h,{\hat\Omega}_h\}_{h=1,...,\hat M}$ are those indicated in Proposition \ref{statchar_theo1}.
\end{proposition}
{
\begin{IEEEproof}
The CDF can be derived from the PDF such as
\begin{align}
F_Z \left( z \right) =\int_0^z f_Z(t)dt.
\end{align}
From Proposition \ref{statchar_theo1}, we have
\begin{align}
F_Z \left( z \right)=\sum\limits_{j = 0}^M {\sum\limits_{h = 0}^{\hat M} {C_j \hat C_h } }F_{\Gamma \Gamma } \left( {z;\left\{ {\Omega _j ,m_j } \right\};\left\{ {\hat \Omega _h ,\hat m_h } \right\}} \right)
\end{align}
where $F_{\Gamma \Gamma }(\cdot)$ is given by equation (\ref{PhuTuan}).
\end{IEEEproof}

Thus, the CDF of the $\mathcal{P}$-distribution is also given in a simple closed-form. This expression has important relevance in practice, since the CDF of the product of INID Rician RVs has a very complicated form, which involves a double infinite sum of Meijer $G$-functions \cite[eq.~(23)]{Bekkali2015}. Here, the $\mathcal{P}$-distribution function, which includes the Rician shadowed product and the Rician product CDFs as special cases, is only given in terms of finite sums of modified Bessel functions of the second kind. Moreover, this expression also simplifies the CDF of the product built from independent Rayleigh and Rician RVs \cite{Bekkali2015}. 

Also note that a similar expression can be given for the moment-generating function, as well as the central moments, which are very useful for certain wireless applications. For the sake of attaining a complete statistical characterization of the $\mathcal{P}$-distribution, these expressions are provided in the following propositions.

\begin{proposition}[MGF of the $\mathcal{P}$-distribution as a finite mixture]
\label{statchar_col2}
Let $Z$ be the product of two INID squared $\kappa$-$\mu$ shadowed random variables $X$ and $\hat X$ with means $\bar\gamma$ and $\widetilde\gamma$. The corresponding shape parameters $\kappa$ and $\hat\kappa$ are arbitrary non-negative real numbers and the remainder shape parameters $\mu$ and $m$ for $X$, and $\hat\mu$ and $\hat m$ for $\hat X$ are positive integers. Under these conditions $Z=X\hat X$ has the following MGF
\begin{equation}
\label{statchar_eq_15}
\begin{gathered}
  {\M}_Z \left( s \right) = \sum\limits_{j = 0}^M {\sum\limits_{h = 0}^{\hat M} {C_j \hat C_h } } \frac{{e^{ - \frac{1}{{2s\,\Omega _j \hat \Omega _h}}} }}{{\left( { - s\,\Omega _j \hat \Omega _h} \right)^{\frac{{m_j  + \hat m_h  - 1}}{2}} }} \times	\W_{ - \frac{{m_j  + \hat m_h  - 1}}{2},\frac{{m_j  - \hat m_h }}{2}} \left( { - \frac{1}{{s\,\Omega _j \hat \Omega _h}}} \right)
\end{gathered}
\end{equation}
where $\W_{a,b}(\cdot)$ denotes the Whittaker function \cite[eq. 9.220.4]{Gradstein2007}, which can be expressed in terms of the Tricomi hypergeometric function.
\end{proposition}
\begin{IEEEproof}
From (\ref{statchar_eq_3}), the MGF of a $\Gamma\Gamma$ random variable can be expressed as
\begin{equation}
\label{statchar_eq_16}
\begin{gathered}
  {\M}_{\Gamma \Gamma } \left( s \right) =\frac{2}{{\left( {\Omega\hat{\Omega}} \right)^{\frac{{m  + \hat{m} }}{2}} \Gamma \left( {m } \right)\Gamma \left( \hat{m } \right)}}
 \int_0^\infty  {x^{\frac{{m  + \hat{m} }}
{2} - 1} e^{sx} K_{m  - \hat{m} } \left( {\sqrt {\frac{4 \; x}
{{\Omega \hat{\Omega} }}} } \right)} dx. \hfill \\
\end{gathered}
\end{equation}
Applying \cite[eq. 4.17.37]{Erdelyi1954} to (\ref{statchar_eq_16}) and considering Proposition \ref{statchar_theo1} completes the proof.
\end{IEEEproof}

\begin{proposition}[Central moments of the $\mathcal{P}$-distribution]
\label{statchar_col3}
Let $Z$ be the product of two INID squared $\kappa$-$\mu$ shadowed random variables $X$ and $\hat X$ with means $\bar\gamma$ and $\widetilde\gamma$. The corresponding shape parameters $\kappa$ and $\hat\kappa$ are arbitrary non-negative real numbers and the remainder shape parameters $\mu$ and $m$ for $X$, and $\hat\mu$ and $\hat m$ for $\hat X$ are positive integers. Under these conditions, $Z=X\hat X$ has the following central moments
\begin{equation}
\label{statchar_eq_17}
\begin{gathered}
  E\left[ {Z^n } \right] = \sum\limits_{j = 0}^M {\sum\limits_{h = 0}^{\hat M} {C_j \hat C_h } \frac{\left( {\Omega _j \hat \Omega _h} \right)^n }
{{\Gamma \left( {m_j } \right)\Gamma \left( {\hat m_h } \right)}}}   \left( {m_j  + n - 1} \right)! \left( {\hat m_h  + n - 1} \right)!. \hfill \\
\end{gathered}
\end{equation}

\end{proposition}
\begin{IEEEproof}
See Appendix \ref{appendix:centralmoments}.
\end{IEEEproof}

\section{Applications}
\label{Applications}
{
We here present different applications of the main results. They can be divided into three parts, since they are associated to different types of communication systems. First, we employ the results in the context of wireless powered communications. Then, we focus on backscatter communications, where we define two particular set-ups: 1) RF modulated backscatter, and 2) dyadic backscatter systems. Finally, we exemplify the applicability of the $\mathcal{P}$-distribution in the context of D2D communications.}

\subsection{Wireless Powered Communications}
\label{WPC}
Wireless communication systems have been classically analyzed under the assumption of ideal power availability for transmitting and receiving signals. 
However, in many scenarios such as wireless sensor networks or RFID systems, the autonomy (and therefore performance) of mobile devices is limited in practice by the finite capacity of their batteries. Even though batteries can be replaced or recharged, the cost in time, money and flexibility is not acceptable in many situations, and therefore other alternatives relying on ambient energy harvesting are considered. Besides solar or wind, the use of RF energy is recently being considered as an alternative for the operation of wirelessly powered devices. Specifically, wireless powered communications are a promising solution to overcome such limitations, by using dedicated power beacons (PBs) that wirelessly convey the required energy to the network elements to enable their operation \cite{Bi2016}.

\begin{figure}[t]
\centering
  \includegraphics[width=0.61\columnwidth, trim={1cm 3cm 6cm 2cm}, clip]{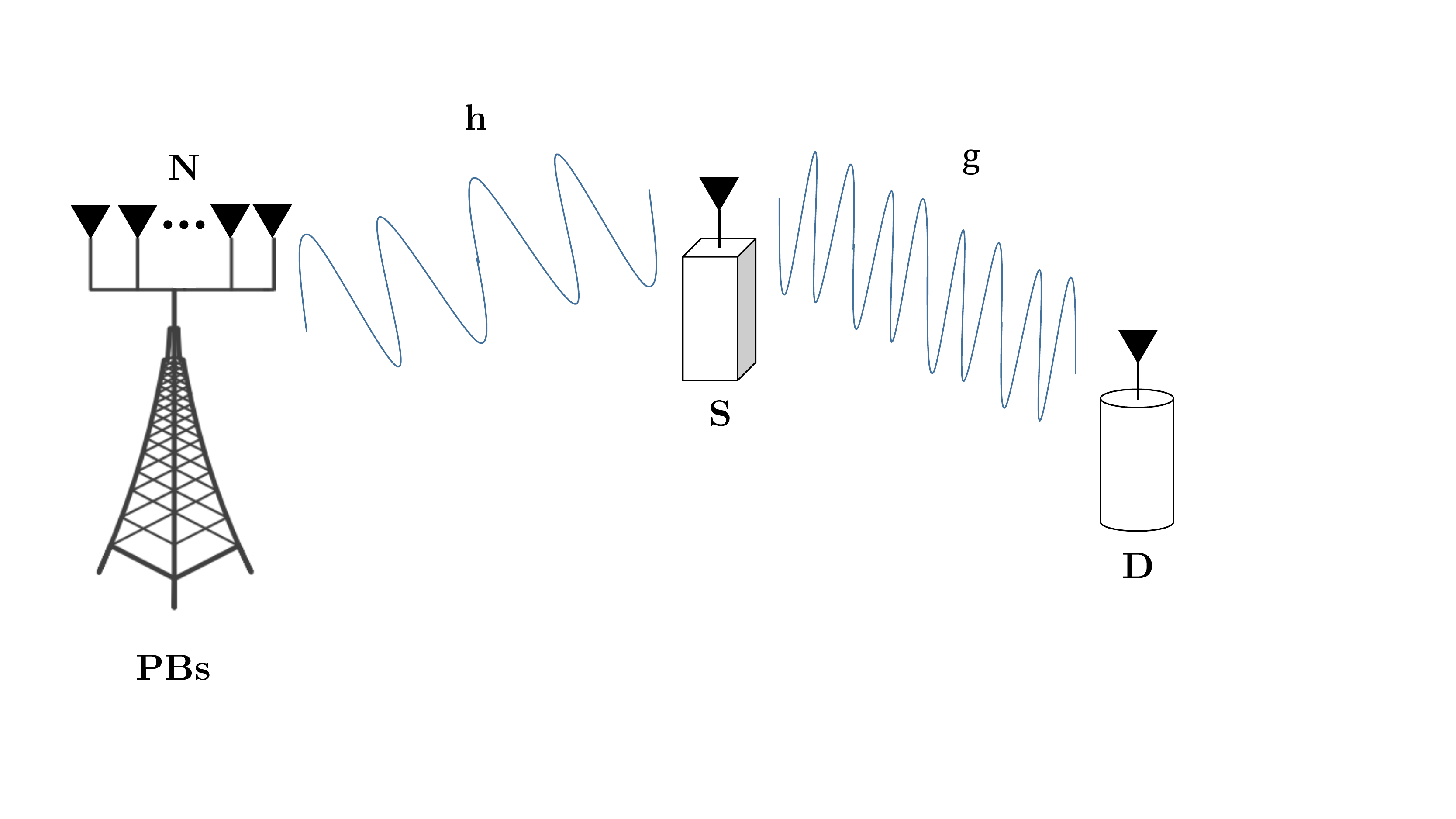}
          \caption{System Model for Wireless Powered Communications.}    
	\label{FigAp2Model}
\end{figure}

Let us consider the scenario in Fig. \ref{FigAp2Model} \cite{Zhong2015,Van2016}, on which a source S communicates with a destination D with the help of dedicated PBs that wirelessly transfer energy to S. Both S and D are equipped with a single antenna, while PBs are equipped with \textit{N} antennas. The system operation follows a \textit{harvest-then-transmit}-like policy for every time transmission interval $T$, as follows: during the first $\tau T$ seconds (with $0<\tau<1$), the source S harvests energy from the PBs. During the energy harvesting phase, the received signal at S can be expressed as
\begin{equation}
\label{Aplication2_eq_1}
y_S  = \sqrt {\frac{P}{{d_1^\alpha  }}} {\bf{hx}}_S  + n_S
\end{equation}
where $P$ is the transmit power at the PB, $d_1$ denotes the distance between PB and S, $\alpha$ is the path loss exponent, ${\bf{h}}$ is a $N$-dimensional row vector and ${\bf{x}}_S$ is a $N$-dimensional column vector, which denote the channel response and the transmitted symbols, respectively, and $n_S$ is the additive white Gaussian noise (AWGN) with variance $N_0$. The elements of ${\bf{h}}=[h_i]$ ($i=1\ldots N$) are assumed to be independent and identically distributed (IID) with unitary variance. Assuming ${\bf{x}}_S$ is formed with optimal beamforming and $n_S$ can be neglected during the harvesting phase, the total energy received at the end of the first phase is \cite[eq.~(5)]{Zhong2015}
\begin{equation}
\label{Aplication2_eq_2}
E_n  = \frac{{\eta \left\| {\bf{h}} \right\|^2 P\tau T}}{{d_1^\alpha  }}
\end{equation}
where $0 < \eta < 1$ is the energy conversion efficiency. 

In the second phase, S transmits information to D using the energy harvested in the first phase during $(1-\tau)T$ seconds. Hence, the received signal $y_D$ at D is given by 
\begin{equation}
\label{Aplication2_eq_3}
y_D  = \sqrt {\frac{{E_n }}{{(1 - \tau )Td_2^\alpha  }}} gs_0  + n_d
\end{equation}
where $d_2$ denotes the distance between S and D, $g$ is the channel response following an arbitrary fading distribution with unit variance, $s_0$ is the information symbol with unit energy, and  $n_d$ is AWGN with variance $N_0$. Therefore, the instantaneous end-to-end signal to noise ratio (SNR) can be computed as 
\begin{equation}
\label{Aplication2_eq_4}
\gamma  = \frac{{\tau \eta \left\| {\bf{h}} \right\|^2 \left| g \right|^2 P}}{{\left( {1 - \tau } \right)d_1^\alpha  d_2^\alpha  N_0 }}.
\end{equation}

Direct inspection of (\ref{Aplication2_eq_4}) reveals that the distribution of $\gamma$ is that of the product of $ \left\| {\bf{h}} \right\|^2$ and $\left| g \right|^2$, which is ultimately related to the distribution of the product of the underlying fading channels between PBs and S, and between S and D. As argued in \cite{Zhong2015}, the link between PBs and S is inherently LOS because of the relatively short distance between both agents. However, the consideration of the Rician distribution to model the small-scale fading in the PBs-S link is associated to a higher mathematical complexity. For this reason, the Rician distribution was approximated by the Nakagami-$m$ distribution in \cite{Zhong2015}, with $m =  (1 + K)^2/(1 + 2 K)$. In turn, the S-D link will be NLOS or LOS depending on the specific set-up: both situations were addressed in \cite{Zhong2015} and \cite{Van2016} by resorting to Rayleigh and Nakagami-$m$ fading, respectively.

In the most general set-up for the system model in Fig. \ref{FigAp2Model}, both the PBs-S and the S-D links are LOS, and therefore the product channel associated to LOS scenarios is the natural choice for characterizing the behavior of the end-to-end SNR. We here propose the use of the $\mathcal{P}$-distribution introduced in Section \ref{statchar} for this application, as a workaround to characterize the distribution of $\gamma$ when the Rician distribution is considered. Because $ \left\| {\bf{h}} \right\|^2$ can be expressed as the sum of $N$ squared Rician random variables (i.e. a $\kappa$-$\mu$ distribution with $\kappa=K$ and $\mu=N$), and assuming $|g|^2$ to be Rician distributed, the distribution of $\gamma$ is that of the product of $\kappa$-$\mu$ and Rician random variables. Thus, it arises as a special case of the $\mathcal{P}$-distribution. Compared to the approximation in \cite{Zhong2015}, our approach has a number of benefits which can be summarized as follows:

\begin{itemize}
\item The Rician shadowed distribution ($\kappa$-$\mu$ shadowed distribution with $\mu=1$) and the Rician distribution have a diversity order equal to one, as opposed to the Nakagami-$m$ distribution, for which the diversity order is $m$. Thus, approximating the Rician distribution by the Rician shadowed distribution does not affect the diversity order. As we will later see, this has an impact on the asymptotic performance for low SNR values.
\item In practice, LOS channels will not be purely Rician because of the inherent fluctuation of the LOS component \cite{Romero2017}. In fact, the $\kappa$-$\mu$ shadowed fading model always provides a better fit to real measurements than the Rician fading model alone, just because the latter is a special case of the former. Thus, the $\mathcal{P}$-distribution is not only simpler and more general than the Rician product distribution, but also closer to the real behavior of the fading channel. 
\end{itemize}

With all the above considerations, the performance of WPC systems in LOS scenarios can be easily evaluated when considering the $\mathcal{P}$-distribution. Assuming that S transmits at a constant rate $R_c$, which may be subjected to outage due to fading, the average throughput can be evaluated as
\begin{equation}
\label{Aplication2_eq_5}
R_{DC}  = \left( {1 - P_{\rm out} } \right)R_c \left( {1 - \tau } \right)
\end{equation}
where $P_{\text{out}} = \Pr \left\{ {\gamma  < \gamma _{\rm th} } \right\}$ is the outage probability, being $\gamma_{\rm th}$ the minimum SNR required for a reliable
communication. As previously stated, the distribution of the product $\left\| {\bf{h}} \right\|^2 \cdot \left| g \right|^2$ can be modeled as a product of two independent squared $\kappa$-$\mu$ shadowed variables with a proper choice of parameters. Thus, the outage probability can be obtained from Proposition 2 as

\begin{equation}
\label{Aplication2_eq_6}
P_{\rm out}  = F_{\gamma} \left( {\frac{{\left( {1 - \tau } \right)d_1^\alpha  d_2^\alpha  N_0 }}{{\tau \eta P}}\gamma _{\rm th} } \right).
\end{equation}

\vspace{2mm}

\subsection{Backscatter Communications}
\label{Backscatter}
Backscatter radio systems base their operation in the ability to detecting power from specular reflections, and can be traced back to the late 40's \cite{Stockman1948}. These systems have been widely used in tropospheric communications or radar, evolving to other applications such as  RFID \cite{Bekkali2015} and RF modulated backscatter (RFMB), also known as modulated radar cross section or sigma modulation \cite{Kim2003}.

Here, we revisit the results provided in \cite{Kim2003} \cite{Bekkali2015} and \cite{Griffin2010}, where backscatter communications are modeled with Rician product channels. We will see that employing the $\mathcal{P}$-distribution simplifies the theoretical analysis, while also improving the accuracy when fitting to real channel measurements.

\subsubsection{RF Modulated Backscatter Systems}
\label{section:RFmodulatedB}
\hfill\\
RFMB, {also known as modulated radar cross
section or sigma modulation}, is an RF technique useful for short range, typically $1-15 {\rm m}$, and low data rate applications, i.e. up to tens of kbps \cite{Finkenzeller2010book}. { Since this technique does not employ amplifiers, service life for system batteries is largely improved. This technique has several applications, including nonstop toll collection, electronic shelf tags, freight container identification and chassis identification in automobile manufacturing \cite{Kim2003}}.

Its operation principle can be illustrated by Fig. \ref{FigAp1Model}. The RF carrier is modulated at the low-power modulator, which gives a reflected signal, also known as backscattered signal. {The exact characterization of such resulting channel involves the product of two LOS RVs, i.e. one can express the received signal power $P_R$ as \cite[eq.~(12)]{Bekkali2015}
\begin{equation}
\label{EqBekkali}
P_R= \bar{P}_R \vert h_f\vert^2 \vert h_b\vert^2
\end{equation}
where $\bar{P}_R$ is average power received by the receiver antenna, while $h_f$ and $h_b$ are the normalized fading coefficients of each link. From (\ref{EqBekkali}), we observe that the resulting channel is equivalent to a cascade of two channels due to the forward and reverse links, so we can use the $\mathcal{P}$-distribution to describe the channel behavior.}

\begin{figure}[t]
\centering
  \includegraphics[width=0.61\columnwidth, trim={3cm 1cm 10cm 1cm}, clip]{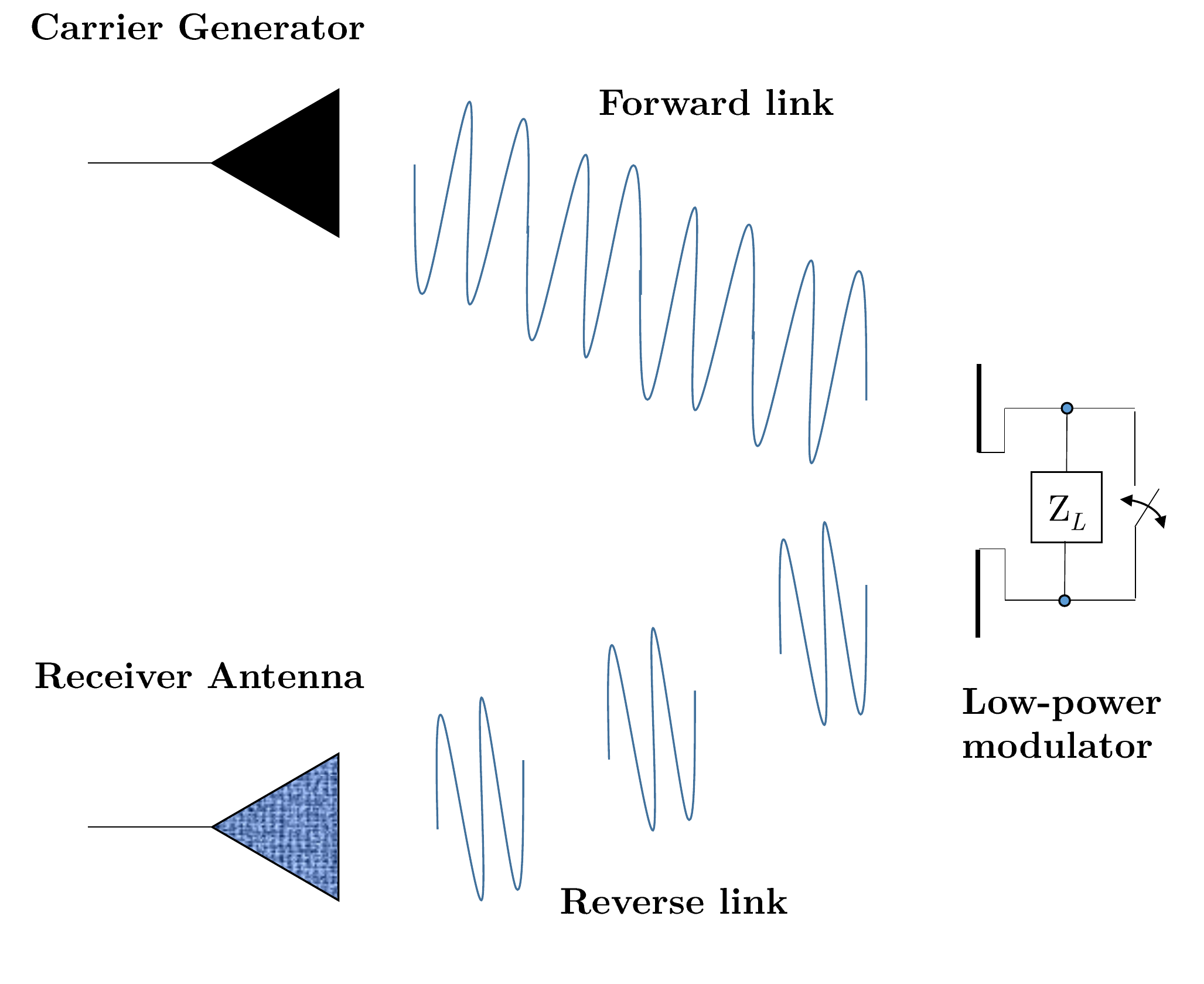}
          \caption{System Model for RFMB.}    
	\label{FigAp1Model}
\end{figure}

\subsubsection{Dyadic Backscatter Systems}
\hfill\\
When compared to the previous backscatter system presented, the so-called dyadic ones employ multiple antennas in both forward an reverse links to reduce the fading severity. The most general set-up comprises an $M$-antenna reader transmitter, $L$ RF tags and an $N$-antenna reader receiver, and the equivalent channel model is usually referred to as the dyadic backscatter channel (DBC) \cite{Griffin2008, Griffin2010}. The use of multiple RF tags effectively reduces the severity of fading in NLOS scenarios, as the equivalent channel can be seen as a product channel built from the sum of products of Rayleigh channels, thus taking advantage of the \textit{pinhole diversity} as $L$ increases \cite{Griffin2008, Griffin2010}.

The theoretical formulation of the DBC is based on the Rayleigh product distribution, mainly due to tractability reasons. However, in these scenarios the forward and reverse links are inherently LOS, as argued before. Thus, a better fit to field measurements is exhibited when considering a DBC built from the product of Rician fading channels.

\begin{figure}[t]
\centering
  \includegraphics[width=0.61\columnwidth, trim={3cm 1cm 10cm 1cm}, clip]{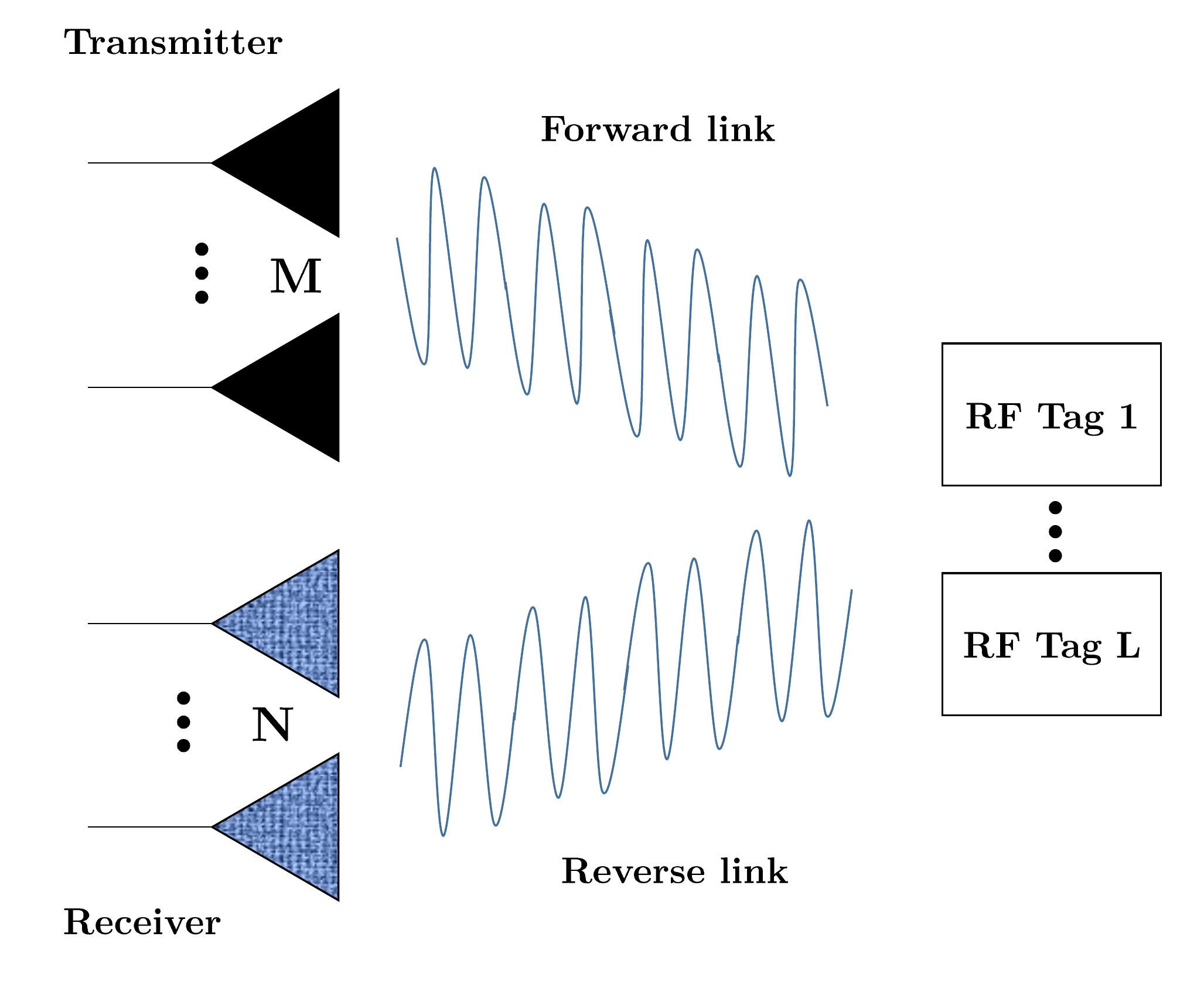}
          \caption{System Model for Dyadic Backscatter Channel.}    
	\label{FigDyatic}
\end{figure}

Let us consider the system model in Fig. \ref{FigDyatic}. The transmitter unit is equipped with $M$ antennas and the receiver unit has $N$ antennas. The resulting channel is the cascade of two channels, due to the forward and backscatter links. We denote it ${M\times L\times N}$ channel. The signal amplitude $y_j$ received at the $j$th antenna is proportional to the sum of $M$ products of the form
\begin{equation}
\label{Aplication1_eq_1}
y_j  \propto \left( {h_1  + h_2  + ... + h_M } \right)g_j 
\end{equation}
where $h_i$ and $g_j$ are the forward and backscatter channel responses from antennas $i$ and $j$, modeled as complex Gaussian circularly symmetric random variables with variances $\sigma_f^2$ and $\sigma_b^2$, respectively. Assuming independency among the $h_i$ elements, the sum  $\left( {h_1  + h_2  + ... + h_M } \right)$ is also a complex Gaussian variable but of variance $M \sigma_f^2$. The effect of using $M$ antennas translates into a scaling of the overall power received at the $j$-th antenna port by $M$. Thus, $y_j$ is proportional to a product of two complex Gaussian variables. In a general LOS scenario, the signal envelope $\left |y_j \right|$ will be the product of two Rician variables. As in the previous sections, we can characterize such product with the $\mathcal{P}$-distribution.

\subsection{D2D communications}
\label{D2D}
Very recently, a composite $\kappa$-$\mu$ fading model was used in \cite{Bhargav2018} in the context of D2D communications operating in the 5.8 GHz band. Different indoor and outdoor scenarios covering LOS and NLOS conditions were considered, and the composite $\kappa$-$\mu$ fading model was shown to provide a good fit to field measurements\footnote{Details on the specific measurement set-up can be found in \cite{Bhargav2018}, and are not reproduced here for the sake of compactness.}. However, as the authors in \cite{Bhargav2018} point out, some counterintuitive results were obtained for the cases on which LOS propagation was considered. Specifically, the value for the parameter $\kappa$ was lower than one despite the clear LOS set-up. A plausible explanation for this was that the LOS was affected by human-body shadowing caused by the random movements of the test subjects, causing a partial obstruction of the dominant signal component.
Motivated by the ability of the $\mathcal{P}$-distribution to model this sort of physical effect, we will later show how \emph{both} the LOS condition of the link \emph{and} the fluctuation on the LOS component are captured by our model, while also providing a good fit to field measurements.

\section{Numerical Results}

{
This section contains all the numerical results regarding the wireless communication applications presented in the previous section. For the sake of clarity, we have mirrored the structure of the previous section, i.e. we have divided this section in three main parts corresponding to the wireless powered, backscatter and D2D communication numerical results.}

\label{NUM}
\subsection{Wireless Powered Communications}

We now use the analytical expressions in Section \ref{WPC} to study the system performance of WPC systems operating in LOS conditions. Our aim is to evaluate to what extent the classical approximation of the Rician distribution through a Nakagami-$m$ distribution \cite{Nakagami1960,AlouiniBook} can be used in the context of WPC scenarios.{ For the sake of comparison and in coherence with those used in \cite{Zhong2015}, the following set of parameters is considered}: $R_c = 1$ bps/Hz, which implies an outage SNR threshold given by $\gamma_ {\rm th} = 2^{R_c} - 1 = 1$, the harvesting time is set to $50\%$ of the interval $T$, the energy conversion efficiency is set to $\eta = 0.4$, the path loss exponent is set to $\alpha = 2.5$, and distances are set to be $d_1 = 8$ m and $d_2 = 15$ m, respectively. 

In Fig. \ref{FigAp2_1}, the throughput obtained from (\ref{Aplication2_eq_5})  is evaluated as a function of the average SNR, for different numbers of antennas at the PBs. Because of the beamforming strategy used by the PBs, the distribution of $\left\| {\bf{h}} \right\|^2$ is that of a squared $\kappa$-$\mu$ random variable. We first assume that the channel between the source S and the destination D is NLOS as in \cite{Zhong2015}, so it can be safely modeled by a Rayleigh fading channel. Thus, we here compare two alternatives for evaluating ({\ref{Aplication2_eq_5}}): the first one is approximating $\left\| {\bf{h}} \right\|^2$ by a squared Nakagami-$m$ (gamma) distribution with $m =  (1 + K)^2/(1 + 2 K) \cdot N$ as in \cite{Zhong2015}, and then using the statistics of a Nakagami-Rayleigh product channel; the second one is using the $\mathcal{P}$-distribution with $\kappa=K$, $\mu=N$, $\hat\kappa=0$, $\hat\mu=1$, and sufficiently large $m$ and $\hat{m}$ (i.e. $m=\hat m = 20$). We consider $K=3+\sqrt{12}$ as in \cite{Zhong2015}. We observe that both approaches yield very similar results for the set of parameters here considered. Thus, the approximation can be safely used when considering a product channel built from a LOS and a NLOS individual channels, for the evaluation of (\ref{Aplication2_eq_5}) { for the whole range of $P/N_0$ values here considered.}

\begin{figure}[t]
\centering
  \includegraphics[width=0.61\columnwidth]{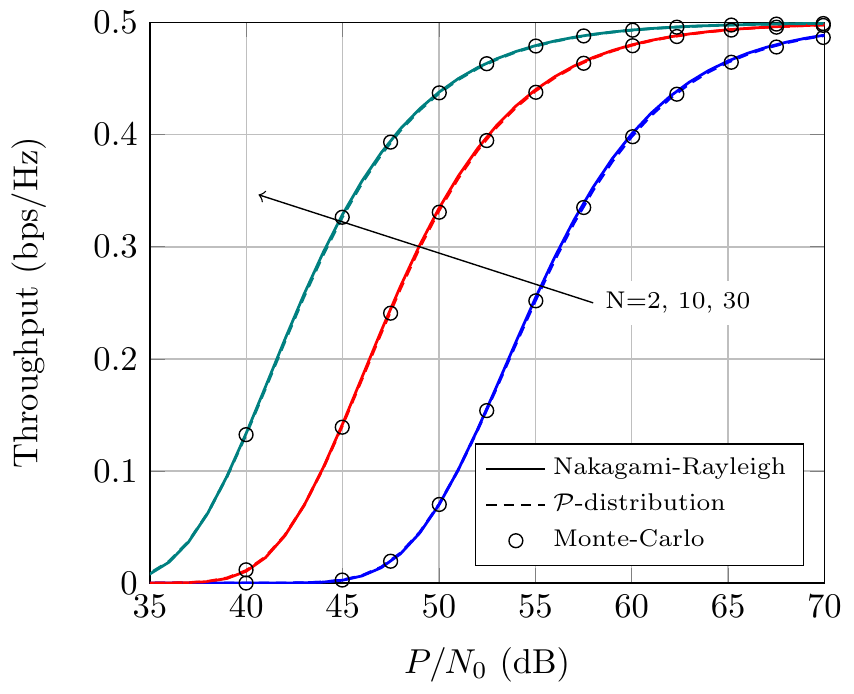}
          \caption{Average Throughput $R_{DC}$ vs. average SNR, for different values of $N$. LOS$\times$NLOS scenario. MC simulations correspond to the Rician-Rayleigh case.}    
	\label{FigAp2_1}
\end{figure}

\begin{figure}[t]
\centering
  \includegraphics[width=0.61\columnwidth]{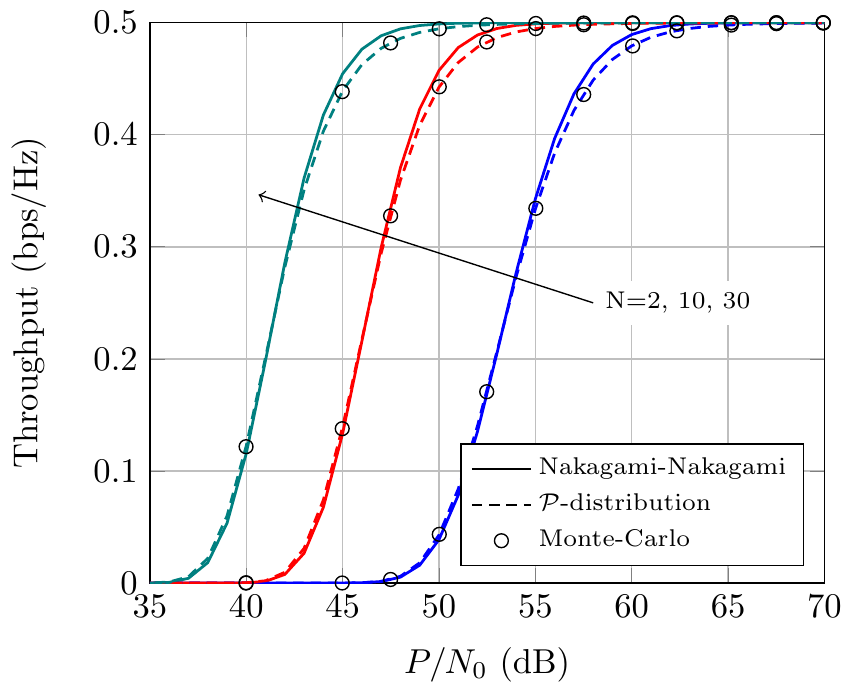}
          \caption{Average Throughput $R_{DC}$ vs. average SNR, for different values of $N$. LOS$\times$LOS scenario. MC simulations correspond to the Rician-Rician case.}
	\label{FigAp2_2}
\end{figure}

However, things change when \textit{both} channels are considered to be LOS. In Fig. \ref{FigAp2_2}, the S-D link is also assumed to be LOS with equal $K$ parameter as in Fig. \ref{FigAp2_1}. {For higher values of $P/N_0$}, we now see a noticeable difference between the approximation in \cite{Zhong2015} and the exact result using the $\mathcal{P}$-distribution with $\kappa=\hat\kappa=K$, $\mu=N$, $\hat{\mu}=1$ and $m=\hat m = 20$, which perfectly matches the Monte-Carlo simulations run for the Rician product case. This becomes even more evident when analyzing the asymptotic behavior of the outage probability in Fig. \ref{FigAp2_3}, where the different diversity orders of the Nakagami-$m$ distribution and the Rician distribution can be observed. Thus, it is not recommended to use the Nakagami-$m$ product channel as an approximation of the Rician product channel, due to the lack of accuracy when modeling LOSxLOS channels, especially in the high-SNR regime.

\begin{figure}[t]
\centering
  \includegraphics[width=0.61\columnwidth]{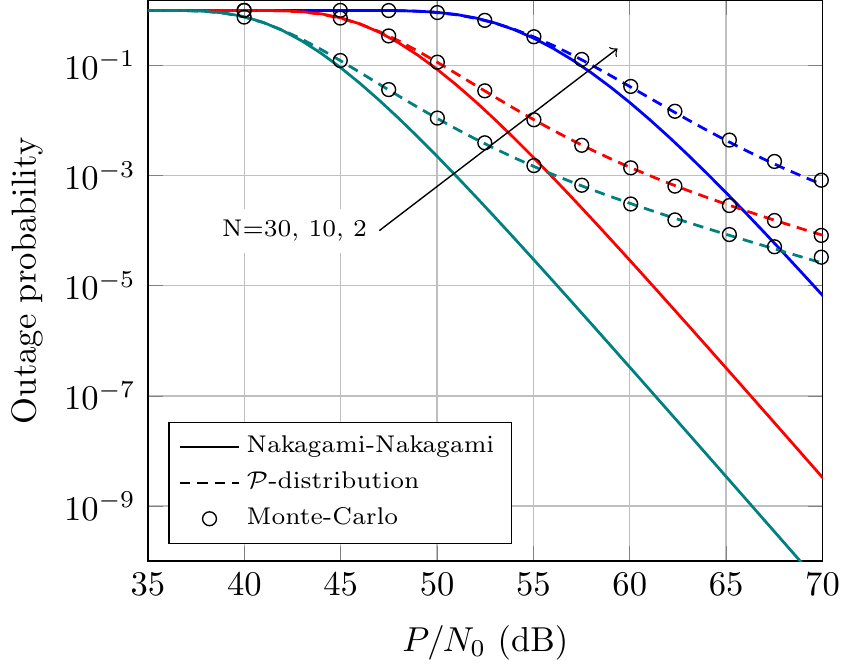}
          \caption{Outage probability vs. average SNR, for different values of $N$. LOS$\times$LOS scenario. MC simulations correspond to the Rician-Rician case.}
	\label{FigAp2_3}
\end{figure}

\subsection{Backscatter Communications}

{
We here employ the theoretical expressions from Section \ref{Backscatter} to exemplify the advantages of our model with respect to other approaches available in the literature for different backscatter communication systems. We will discuss the accuracy of the proposed model, as well as we will quantify its most relevant characteristics, with a special focus on its flexibility for fitting purposes.}

\subsubsection{RF Modulated Backscatter Systems}
\hfill\\
{
We here consider the scenario in \cite{Kim2003}, which has been described in Section \ref{section:RFmodulatedB}.} Like \cite{Kim2003}, we assume that both forward and reverse links have the same statistics. Fig. \ref{FigKimresults} shows a comparison between the CDF measurement performed in \cite{Kim2003}, the Rician-Rician fitting provided in \cite{Kim2003} and the corresponding theoretical curve obtained with the $\mathcal{P}$-distribution. In addition to simplifying the characterization of such product LOS channel, we highlight that there is an outstanding agreement between measurements and the theoretical analysis with the $\mathcal{P}$-distribution. Although the Rician model is theoretically obtained by tending $m$ to infinity, {we obtain the best fit with a moderate value of the shadowing parameter ($m=4$) and a slightly larger $\kappa$. This is justified in Appendix \ref{apC}}, where we see that the CDF under $\kappa$-$\mu$ fading is almost identical to the $\kappa$-$\mu$ shadowed fading CDF with finite $m>\mu$ and larger $\kappa$. By only evaluating $64$ ($3$ nested finite sums of $4$ terms) modified Bessel of second kind, which is immediate, we see a slightly better fitting with our $\mathcal{P}$-distribution than the corresponding one with the Rician product model in \cite{Kim2003}. To evaluate the improvement, we define an error factor $\epsilon$ based on a modified version of the Kolmogorov-Smirnov (KS) statistic, i.e.
\begin{align}
\label{KS}
\epsilon\triangleq\max_{x}\vert\log_{10} \widetilde{F}_Z(x)-\log_{10}F_Z(x) \vert
\end{align}
where $\widetilde{F}_Z(x)$ and $F_Z(x)$ are the empirical and theoretical CDFs, respectively. Note that, with the above definition, a value of $\epsilon=1$ corresponds to a difference of one order of magnitude between empirical and theoretical CDFs. 

The procedure followed for the fitting is described as follows: we bound the maximum value of $m$ to be considered in the fitting ($m\leq30$), and then we find the values of $\kappa$ and $\mu$ that minimize the KS statistic for a given integer $m$. We finally choose the set of parameters $\kappa$, $\mu$ and $m$ that achieve a minimal value of the KS statistic, which in this case are $\kappa=2.6$ and $m=4$, with $\mu=1$ as in the Rician case.
With such KS statistic, the error factor value for the fitting proposed in \cite{Kim2003} with the Rician product model is $\epsilon^{\rm Rician}=0.0882$, while the corresponding one with the $\mathcal{P}$-distribution is $\epsilon^{\mathcal{P}}=0.0793$. 

Therefore, the advantages of using the $\mathcal{P}$-distribution instead of the classical Rician product distribution are evident. The analysis of systems that involve two different communication LOS links is simplified, which facilitates the computation of relevant performance measures, and, at the same time, using the $\mathcal{P}$-distribution does not compromise the accuracy when fitting measurements, but it rather improves it.

\begin{figure}[t]
\centering
  \includegraphics[width=0.61\columnwidth]{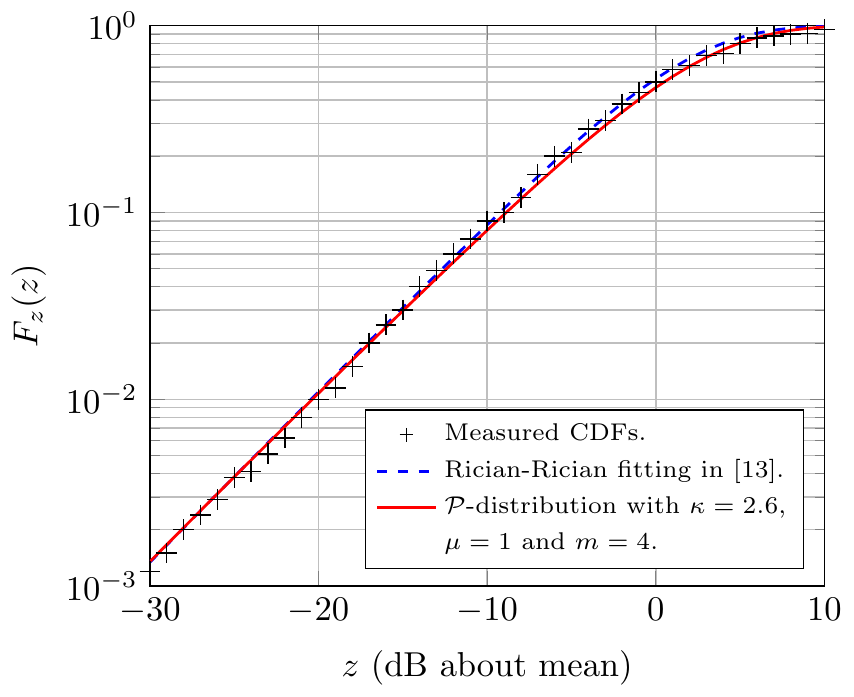}
          \caption{Empirical vs. theoretical CDFs of RF modulated backscatter system. { Rician-Rician fitting with $K=1.90$ (i.e. $K=2.8$dB) in \cite{Kim2003}.}}    
	\label{FigKimresults}
\end{figure}

\subsubsection{Dyadic Backscatter Systems}
\hfill\\
{Three different DBCs are here considered: $1\times 1 \times 1$, $1\times 2\times 1$ and $1\times 2\times 2$ configurations, i.e. we consider single-tag (STAG) and double-tag (DTAG) configurations, with a single antenna at the transmitter side, and one or two antennas at the receiver side. There are two sets of measurements, which correspond to the different configurations for the STAG and DTAG cases. Using the original notation in \cite{Griffin2010}, they are referred to as Rx1, Rx2 and MRC (since maximal ratio combining is performed in the 2-receive antenna configuration) in Fig. \ref{Fig8a} and Fig. \ref{Fig8b}, respectively. Further details on the specific measurement set-up can be found in \cite{Griffin2010}.

Figs. \ref{Fig8a} and \ref{Fig8b} present the empirical results for the DBC provided in \cite[Fig. 8a]{Griffin2010} and \cite[Fig. 8b]{Griffin2010}. They also depict two different fittings with the Rician product {(i.e. $m\rightarrow\infty$)} and the $\mathcal{P}$ distributions. Specifically, the empirical CDFs from measured data are represented using dotted lines, whereas the Rician product and the $\mathcal{P}$ distribution fittings are represented using dashed and solid lines, respectively. Different colors have been used to identify the CDFs at Rx1, Rx2 and after MRC. We see that the empirical CDFs are well modeled with both LOS product models. To quantify the goodness of each fitting, we have computed the KS error factor defined in (\ref{KS}). \mbox{Table \ref{table_2}} presents the values of this error factor for each case, as well as the distribution parameters used for the fitting in each case. In general terms, we observe that the simpler $\mathcal{P}$-distribution always provides a better fit than the Rician product distribution. In some cases, e.g, in the DTAG scenario in Fig. \ref{Fig8b}, we see that the fitted CDFs are practically overlapped for the single antenna configurations. However, the KS error factor obtained for the $\mathcal{P}$-distribution is always better than that of the Rician product case, especially in the STAG scenario. We also see that an improved fit is obtained for the dual-antenna configuration using MRC when considering the $\mathcal{P}$-distribution with $\hat{\mu}=2$. This is coherent with the underlying physical set-up, as using two receive antennas with MRC is equivalent to considering two clusters, which is naturally captured by the $\mathcal{P}$-distribution. We also see that for the DTAG configuration, the use of two tags is not translated into a value of $\mu=2$ in the equivalent channel. This is also explained by the fact that in the presence of a LOS component larger than $3$dB, the pinhole diversity attained by having multiple RF tags is reduced due to correlation and hence the distribution of the equivalent channel is well-approximated by that of a single RF tag \cite{griffin2011fading}. Hence, the use of the $\mathcal{P}$-distribution renders a reduced mathematical complexity, together with a better fit to measurements and an improved physical interpretation.
}

\begin{table}[!t]
\caption{{Parameter values and KS error factor $\epsilon$ for Figs. \ref{Fig8a} and \ref{Fig8b}}}.
\label{table_2}
\centering
\vspace{-8mm}
\begin{tabular}{|c|c|c|c|c|c|c|} 
\hline
\multirow{2}{*}{} & \multicolumn{3}{c|}{STAG} & \multicolumn{3}{c|}{DTAG}  \\ 
\cline{2-7}
                  & Rx1 & Rx2 & MRC           & Rx1 & Rx2 & MRC            \\ 
\hline
Rician      & $K=7,$ $\hat{K}=7$    &  $K=5,$ $\hat{K}=4.9$   &      $K=8.1,$ $\hat{K}=8.2$         & $K=8,$ $\hat{K}=7.9$    &  $K=7.9,$ $\hat{K}=7.9$   &     $K=11,$ $\hat{K}=15$           \\ 
\hline
$\mathcal{P}$           & $\kappa=12,$ $\hat{\kappa}=12.1$     &   $\kappa=8,$ $\hat{\kappa}=8.1$  &     $\kappa=10.1,$ $\hat{\kappa}=12$          & $\kappa=10,$ $\hat{\kappa}=9.9$    &  $\kappa=11,$ $\hat{\kappa}=11$   &    $\kappa=12,$ $\hat{\kappa}=15$            \\ 
  & $\mu=1,$ $\hat{\mu}=1$     &   $\mu=1,$ $\hat{\mu}=1$    &     $\mu=1,$ $\hat{\mu}=2$            &  $\mu=1,$ $\hat{\mu}=1$     &   $\mu=1,$ $\hat{\mu}=1$    &   $\mu=1,$ $\hat{\mu}=2$        \\
 & $m=9,$ $\hat{m}=9$     &  $m=8,$ $\hat{m}=8$     &       $m=8,$ $\hat{m}=10$          &   $m=20,$ $\hat{m}=21$    &  $m=15,$ $\hat{m}=15$     &     $m=30,$ $\hat{m}=20$     \\
\hline
   {$\epsilon^{\rm Rician}$}               &   0.1135  & 0.3848    &  0.3255    &  0.2107   &  0.2589   &   0.3282             \\ 
\hline
  {$\epsilon^{\mathcal{P}}$}                 &  0.1067    & 0.3542    &    0.3172    &   0.2086    &  0.2555   &    0.2660           \\
\hline
\end{tabular}
\end{table}

{
\begin{figure}[t]
\centering
  \includegraphics[width=0.61\columnwidth]{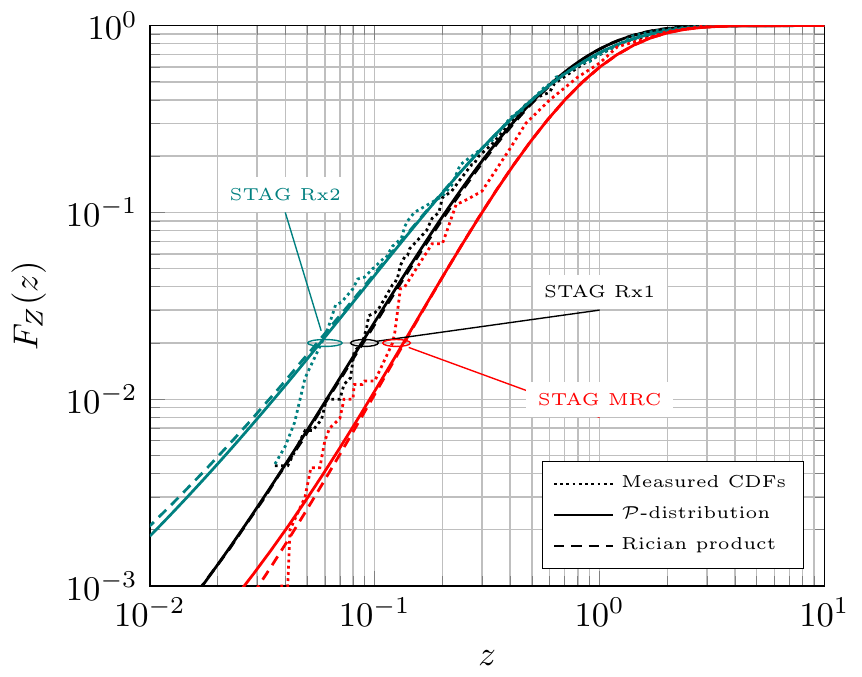}
            \caption{{Empirical vs. theoretical CDFs in {\cite[Fig. 8a]{Griffin2010}} for a STAG configuration. Colored lines indicate the CDFs at Rx1 (black), Rx2 (teal) and after MRC (red). Parameter values for the fitting with Rician product and $\mathcal{P}$-distribution are summarized in Table \ref{table_2}.}} 
          \label{Fig8a}
\end{figure}

\begin{figure}[t]
\centering
  \includegraphics[width=0.61\columnwidth]{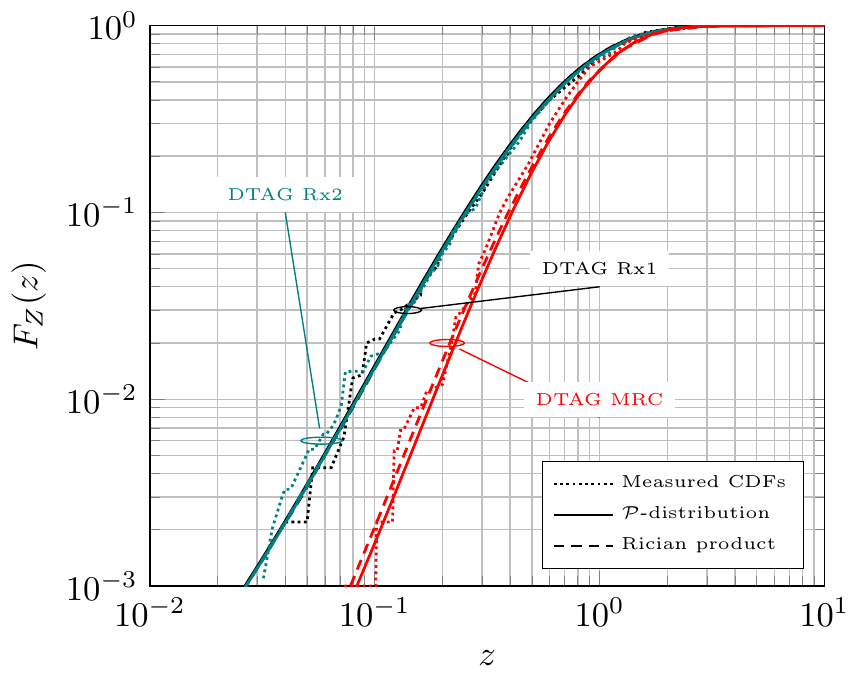}
                    \caption{{Empirical vs. theoretical CDFs in {\cite[Fig. 8b]{Griffin2010}} for a DTAG configuration. Colored lines indicate the CDFs at Rx1 (black), Rx2 (teal) and after MRC (red). Parameter values for the fitting with Rician product and $\mathcal{P}$-distribution are summarized in Table \ref{table_2}.}}
          \label{Fig8b}
\end{figure}
}
\clearpage

\subsection{D2D communications}

We use the set of field measurements in \cite{Bhargav2018} for the indoor and outdoor LOS configurations, in order to determine whether the $\mathcal{P}$-distribution fits well empirical data while allowing a plausible physical interpretation of the channel fading parameters. As suggested in \cite{Bhargav2018} from the observation of the fitting procedure, a single cluster of multipath waves is enough to characterize this component both in indoor and outdoor environments. Hence, and in coherence with the physical models of the $\kappa$-$\mu$ and $\kappa$-$\mu$ shadowed distributions (which assume an integer number of clusters), we fix the parameters $\mu=\hat\mu=1$. Note that in this situation, both the composite $\kappa$-$\mu$ and the $\mathcal{P}$-distribution have four shape parameters, so that the use of the latter does not grant any benefits from having extra parameters.

Following the procedure described in \cite{Bhargav2018}, the composite fading model is built as the product of two independent RVs for which the receive signal envelope $r_{\rm comp}=r\cdot\hat r$, where it is assumed without loss of generality that $\mathbb{E}\{r\}=1$, with $\mathbb{E}\{\hat r\}=\tilde r$. We obtain the envelope PDF of the $\mathcal{P}$-distribution by using a simple transformation of random variables. Using the empirical PDFs given in \cite{Bhargav2018}, we obtain the optimal parameter estimates for the $\mathcal{P}$-distribution that minimize the mean-square error (MSE) between the empirical and target PDF. These are summarized in Table \ref{tableAjuste} and Fig. \ref{FigAjuste1}, and compared to those obtained using the composite $\kappa$-$\mu$ fading model. 

\begin{table*}[t]
\centering
{
\caption{Parameter estimates for the $\mathcal{P}$-distribution vs. the composite $\kappa$-$\mu$ distribution fitted to measured data.}
\label{tableAjuste}
\begin{tabular}{|c|c|c|c|c|c|c|c|c|c|c|c|}
\hline\hline
Channel type & $\kappa$ & $\mu$& $m$ & $\hat \kappa$ & $\hat \mu$& $\hat m$ & $\tilde r$ &$\rm MSE(\%)$ \\ 
\hline 
Indoor LOS  $\mathcal{P}$      & 4.79    & 1 & 10 & 5.25   & 1& 1& 0.90 & 1.02\%\\ 
Indoor LOS  \cite{Bhargav2018}      & 3.94    & 0.67 & $\infty$  & 0.72         & 1.18& $\infty$ &0.89& 1.23\%\\ 
\hline 
Outdoor LOS  $\mathcal{P}$        & 1.87    & 1& 19  & 1.67   & 1       & 6 & 0.93& 1.66\%\\ 
Outdoor LOS \cite{Bhargav2018}        & 1.41    & 1.08  &$\infty$ &     1.00     & 1.14 & $\infty$ & 0.93 &1.94\%\\ 
\hline 
\end{tabular}
}
\end{table*}

\begin{figure}[t]
\centering
  \includegraphics[width=0.61\columnwidth]{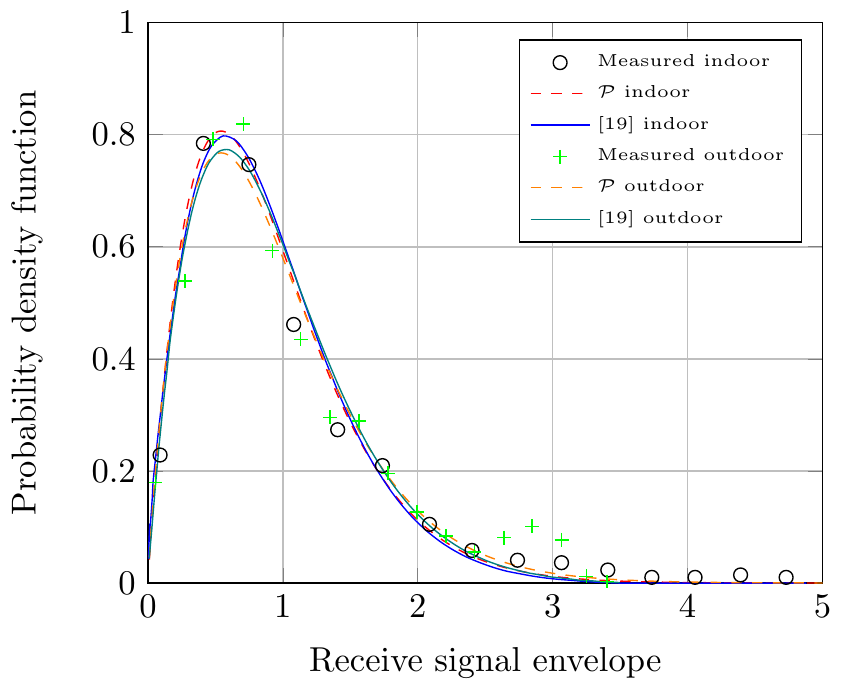}
          \caption{Empirical and theoretical PDFs of the $\mathcal{P}$-distribution fitted to the D2D channel measurements in \cite{Bhargav2018} for LOS
indoor and outdoor environments. The receive signal envelope is normalized the the sample global mean.}
	\label{FigAjuste1}
\end{figure}

Inspection of Table \ref{tableAjuste} reveals important insights from a physical perspective. First, the use of the $\mathcal{P}$-distribution effectively captures the severe fluctuation on the LOS component in the indoor set-up. We see that a parameter $m=1$ is obtained with $\hat\kappa>1$, which is in coherence with a LOS scenario with severe LOS shadowing. Neglecting the effect of the LOS fluctuation causes that the parameter $\hat\kappa$ is underestimated by the composite $\kappa$-$\mu$ fading model. For the rest of scenarios, we see that a larger value of $\kappa$ and $\hat\kappa$ is obtained compared to the deterministic LOS case, which is compensated by a finite value of $m$ and $\hat m$ (i.e., a mild LOS fluctuation). Fig. \ref{FigAjuste1} shows that in both cases, the fitting to empirical data yields rather similar PDFs. However, the use of the $\mathcal{P}$-distribution is more consistent with the underlying physical model (LOS with random fluctuations and a single cluster), and also simpler from an analytical perspective.

\section{Conclusions}
\label{Conclusions}

We have simplified the study of LOS product fading channels, for which results are generally complicated thus far due to the significant analytical challenge posed by the product of Rician RVs.  We have introduced a new model based on the product of two INID $\kappa$-$\mu$ shadowed RVs, which has allowed us to characterize LOS product channels models with very simple closed-form expressions. In particular, we have characterized the PDF, CDF and MGF in terms of finite sums of well-known special functions, which can be found in commercial mathematical packages. The usefulness of the results have been exemplified through the analysis of WPC, backscatter and D2D communication systems. Specifically, we have observed that previous channel approximations based on the Nakagami-$m$ distribution failed to provide good accuracy for such non-ergodic measures in LOSxLOS scenarios. Moreover, in addition to simplifying previous exact theoretical results for \mbox{LOSxLOS} and LOSxNLOS product channels, our model has better flexibility when fitting experimental results, which makes the \mbox{$\mathcal{P}$-distribution} the most suitable choice to model LOS product channels.

\appendices

\section{Proof of Proposition \ref{statchar_theo1}}
\label{appendix:pdf}
The MGF of $Z$ can be computed as
\begin{equation}
\label{statchar_eq_6}
\begin{gathered}
  {\M}_Z \left( s \right) \triangleq E\left[ {e^{X\hat Xs} } \right] =  \hfill \\
  \int_0^\infty  {E\left[ {e^{x\hat Xs} \left| {X = x} \right.} \right]f_X \left( x \right)dx}
  = \int_0^\infty  {f_X \left( x \right){\M}_{\hat X} \left( {sx} \right)dx}.  \hfill \\
\end{gathered}
\end{equation}
Now, according to Lemma \ref{statchar_lema1} the MGF of $\hat X$ in (\ref{statchar_eq_6}) can be expressed in terms of the squared Nakagami-$m$ MGF ${\M}_{\K}$ as follows
\begin{equation}
\label{statchar_eq_7} 
{\M}_{\hat X} \left( sx \right) =\sum\limits_{h = 0}^{\hat M} {\hat C_h {\M}_{\K} \left( {\hat\Omega _h ;\hat m_h ;sx} \right)}.
\end{equation}
Using (\ref{statchar_eq_1}) and (\ref{statchar_eq_7}) in (\ref{statchar_eq_6}) and expanding the integrand yields
\begin{equation}
\label{statchar_eq_8}
\begin{gathered}
  {\M}_Z \left( s \right) = \sum\limits_{j = 0}^M {\sum\limits_{h = 0}^{\hat M} {C_j \hat C_h }  \times }  \hfill \\
  \quad \quad \quad
  \underbrace{\int_0^\infty  {f_{\K} \left( {\Omega _j ;m_j ;x} \right){\M}_{\K} \left( {\hat \Omega _h ;\hat m_h ;sx} \right)dx}}
  _{{\M}_{\Gamma\Gamma} \left( {s;\left\{ {\Omega _j ,m_j } \right\};\{ {\hat \Omega _h ,\hat m_h } \}} \right)}  \hfill \\
\end{gathered}
\end{equation}
where ${{\M}_{\Gamma\Gamma} \left( {s;\left\{ {\Omega _j ,m_j } \right\};\{ {\hat \Omega _h ,\hat m_h } \}} \right)}$ is given in (\ref{statchar_eq_16}). Thus, applying the inverse Laplace Transform in (\ref{statchar_eq_8})
and considering (\ref{statchar_eq_3}) completes the proof.

\section{Proof of Proposition \ref{statchar_col3}}
\label{appendix:centralmoments}
The central moments of the product of two INID squared Nakagami-$m$ random variables are given by
\begin{equation}
\label{statchar_eq_18}
\begin{gathered}
 E\left[ {Z_{\Gamma \Gamma }^n } \right] = \frac{2}{{\left( {\Omega \hat{\Omega}} \right)^{\frac{{m  + \hat{m} }}{2}} \Gamma \left( {m } \right)\Gamma \left( \hat{m } \right)}} \times \\
{\rm{  }}\int_0^\infty  {x^{\frac{{2n + m  + \hat{m} }}{2} - 1} K_{m  - \hat{m} } \left( {\sqrt {\frac{{4x}}{{\Omega \hat{\Omega} }}} } \right)dx} 
. \hfill \\
\end{gathered}
\end{equation}
{ In order to solve the integral in ($\ref{statchar_eq_18}$),} let us consider the following function
\begin{equation}
\label{statchar_eq_10}
\Lambda \left( t \right) \triangleq \int {t^{2q + p - 1} } K_p \left( t \right)dt,
\end{equation}
where $q\geq 1$ and $p\geq 0$ are integer numbers. Taking into account that \cite[eq. 11.3.27]{abramowitz1964}
\begin{equation}
\label{statchar_eq_11}
\frac{d}
{{dz}}z^\nu  K_\nu  \left( z \right) =  - z^\nu  K_{\nu  - 1} \left( z \right),\qquad(\nu>0)
\end{equation}
and iteratively integrating by parts in (\ref{statchar_eq_10}), the following formula is obtained for $\Lambda \left( t \right)$
\begin{equation}
\label{statchar_eq_12}
\Lambda \left( t \right) =  - \sum\limits_{r = 1}^m {2^{r - 1} \frac{{\left( {q - 1} \right)!}}
{{\left( {q - r} \right)!}}t^{2\left( {q - r} \right)} t^{p + r} K_{p + r} \left( t \right)} .
\end{equation}
{
Using \cite[eq. 9.7.2]{abramowitz1964}, it follows that $\Lambda \left( \infty  \right)=0$. On the other hand,} after considering \cite[eq. 9.6.8]{abramowitz1964} and \cite[eq. 9.6.9]{abramowitz1964}, the value for $\Lambda \left( t \right)$ in $t=0$ is given by
\begin{equation}
\label{statchar_eq_13}
\Lambda \left( 0 \right) =  - 2^{q - 1} \left( {q - 1} \right)!\left( {p + q - 1} \right)!2^{p + q - 1}.
\end{equation}

Since $K_{\nu}=K_{-\nu}$, without loss of generality we can consider $m\geq \hat{m}$ in the integral of  (\ref{statchar_eq_18}) and we can work with $K_{|\nu|}$ instead of $K_{\nu}$.
Thus, after the change of variable $
{\frac{{4}}
{{\Omega \hat{\Omega} }}x = t^2 },
$ setting $p=|m-\hat{m}|$ and $q=\hat{m}$, we can obtain
\begin{equation}
\label{statchar_eq_14}
E\left[ {Z_{\Gamma \Gamma }^n } \right] = \frac{{\left( {\Omega \hat{\Omega} } \right)^n }}{{\Gamma \left( {m } \right)\Gamma \left( \hat{m } \right)}}\left( {n + m  - 1} \right)!\left( {n + \hat{m}  - 1} \right)!
.
\end{equation}
After considering Proposition \ref{statchar_theo1}, the proof is complete.
%
{
\section{CDF approximations for the $\kappa$-$\mu$ distribution using the $\kappa$-$\mu$ shadowed distribution}
\label{apC}

In this appendix, we exemplify how LOS distributions arising from the $\kappa$-$\mu$ shadowed distributions can be used for approximating the CDF of the $\kappa$-$\mu$ distribution\footnote{For $\mu=1$, this is equivalent to using the Rician shadowed distribution to approximate the Rician distribution.}. For the sake of notational convenience, the fading parameters for the latter will be denoted as $K$ and $\mu$, whereas the fading parameters for the former will be denoted as $\kappa$, $\mu$ and $m$. 

Let us express the asymptotic approximation for the CDF of the $\kappa$-$\mu$ shadowed distribution in \cite[eq. (13)]{Paris2014} using the standard nomenclature in \cite{Wang2003} as
\begin{equation}
\label{eqnew01}
F_{\kappa\mu m}(\gamma)\approx \frac{a_1}{t+1}\left(\frac{\gamma}{\bar\gamma}\right)^{t+1},
\end{equation}
with
\begin{equation}
\label{eqnew02}
a_1=\frac{\mu^{\mu}(1+\kappa)^{\mu}}{\Gamma(\mu+1)}\left(\frac{m}{\kappa\mu+m}\right)^m
\end{equation}
and $t=\mu-1$. Note that the parameter $t$ is related to the slope behavior of the CDF (i.e. the diversity order), whereas the parameter $a_1$ can be regarded as a power offset. Because the $\kappa$-$\mu$ shadowed distribution and the $\kappa$-$\mu$ distribution have the same diversity order for equal $\mu$ \cite{Lopez2017}, an asymptotic approximation for the $\kappa$-$\mu$ distribution will have the same $t$ as in \eqref{eqnew01}. Thus, we can approximate the CDF of the $\kappa$-$\mu$ distribution by letting $m\rightarrow\infty$ in \eqref{eqnew02} yielding
\begin{equation}
\label{eqnew03}
F_{\kappa\mu}(\gamma)\approx \frac{a_2}{t+1}\left(\frac{\gamma}{\bar\gamma}\right)^{t+1},
\end{equation}
with
\begin{equation}
\label{eqnew04}
a_2=\frac{\mu^{\mu}(1+K)^{\mu}e^{-K \mu}}{\Gamma(\mu+1)}.
\end{equation}

Indeed, the approximations in \eqref{eqnew01} and \eqref{eqnew03} are equal for $\kappa=K$ as both distributions converge $\forall \gamma$ \cite{Lopez2017}. However, both approximations are also coincident provided that $a_1=a_2$. This implies that for a given $K$ there are infinite pairs of $\{\kappa,m\}$ that lead to the same asymptotic behavior. Such values can be obtained from \eqref{eqnew02} and \eqref{eqnew04} that must satisfy:
\begin{equation}
e^{-K}(K+1)=(\kappa+1)\left(\frac{m}{\mu\kappa+m}\right)^{m/\mu},
\end{equation}
Note that $K=\kappa$ and $m\rightarrow\infty$ satisfies the previous equation, {as $e\triangleq \lim_{n\rightarrow\infty}\left(1+\frac{1}{n}\right)^n$}. For a finite $m>\mu$, the value of $\kappa$ that achieves the same asymptotic behavior as the $K$-$\mu$ distribution is necessarily $\kappa>K$. This can be interpreted from the underlying physical meaning of the fading parameters as a trade-off between LOS power and LOS fluctuation: by allowing the LOS component to randomly fluctuate, the fading severity is increased. This effect is compensated by rising $\kappa$, in order for both CDFs to asymptotically coincide. { Similarly, neglecting the fluctuation of the LOS component (i.e. $m\rightarrow\infty$) may lead to underestimating the $K$ parameter \cite{Bhargav2018}, when compared to the case of considering a finite $m$.}

The accuracy of this approach is exemplified in Figs. \ref{FigApp01} and \ref{FigApp02}. In general terms, a larger $K$ requires for a larger $m$ in order for both distributions to behave more similarly. However, setting a value of $m=15$ with $\kappa=14.95$ is enough for practically overlapping with the CDF of $\kappa=10$ and $m\rightarrow\infty$. As the LOS power is increased (i.e. a higher $K$), the value of $m$ required for both distributions to coincide is reduced. Note that all CDFs in the figures have the same asymptotic behavior, indicated by the dashed black line curve

\begin{figure}[t]
\centering
  \includegraphics[width=0.61\columnwidth]{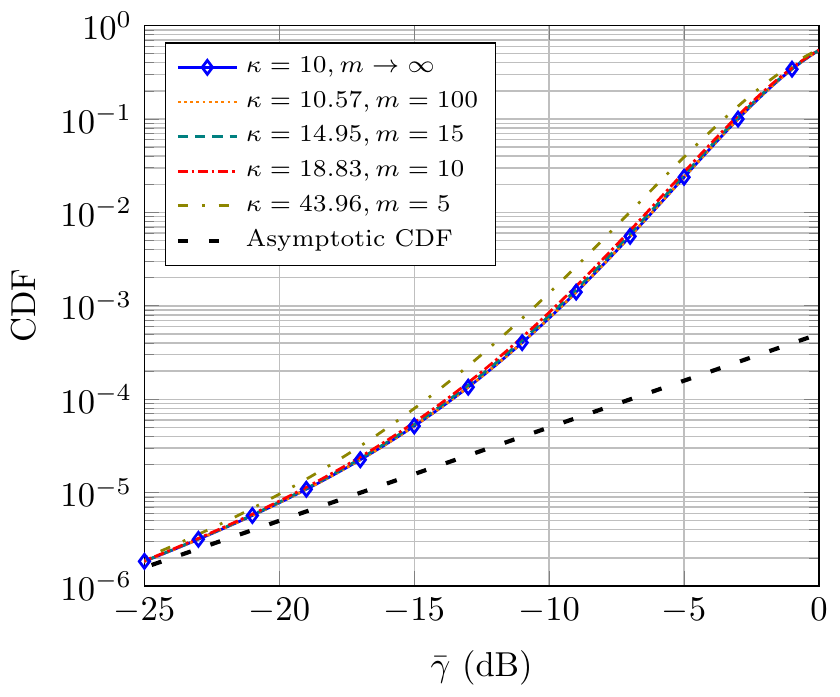}
          \caption{CDF approximations for the $\kappa$-$\mu$ distribution ($\kappa=K=10$, $m\rightarrow\infty)$) using the $\kappa$-$\mu$ shadowed distribution with different pairs of values of $\kappa$ and $m$. Parameter value $\mu=1$. Tail approximation uses \eqref{eqnew01}, or equivalently, \eqref{eqnew03}}.    
	\label{FigApp01}
\end{figure}

\begin{figure}[t]
\centering
  \includegraphics[width=0.61\columnwidth]{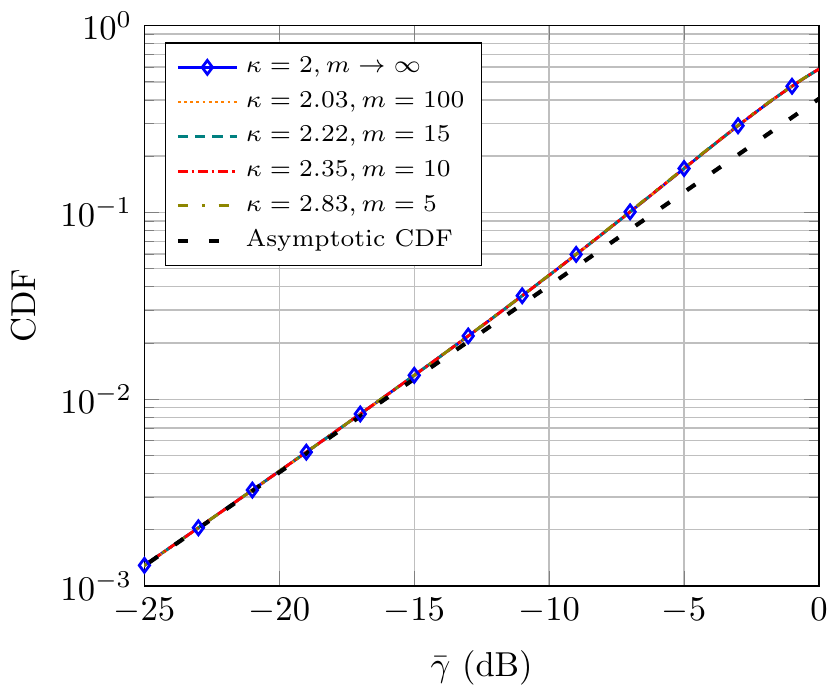}
          \caption{CDF approximations for the $\kappa$-$\mu$ distribution ($\kappa=K=2$, $m\rightarrow\infty)$) using the $\kappa$-$\mu$ shadowed distribution with different pairs of values of $\kappa$ and $m$. Parameter value $\mu=1$. Tail approximation uses \eqref{eqnew01}, or equivalently, \eqref{eqnew03}}.    
	\label{FigApp02}
\end{figure}

This rationale justifies to approximate the distribution of $\kappa$-$\mu$ (or Rician) product channels by means of the distribution of $\kappa$-$\mu$ shadowed (or Rician shadowed) product channels, here referred to as the $\mathcal{P}$-distribution. 
}

\bibliographystyle{IEEEtran}
\bibliography{product3}

\begin{thebibliography}{10}
\providecommand{\url}[1]{#1}
\csname url@samestyle\endcsname
\providecommand{\newblock}{\relax}
\providecommand{\bibinfo}[2]{#2}
\providecommand{\BIBentrySTDinterwordspacing}{\spaceskip=0pt\relax}
\providecommand{\BIBentryALTinterwordstretchfactor}{4}
\providecommand{\BIBentryALTinterwordspacing}{\spaceskip=\fontdimen2\font plus
\BIBentryALTinterwordstretchfactor\fontdimen3\font minus
  \fontdimen4\font\relax}
\providecommand{\BIBforeignlanguage}[2]{{%
\expandafter\ifx\csname l@#1\endcsname\relax
\typeout{** WARNING: IEEEtran.bst: No hyphenation pattern has been}%
\typeout{** loaded for the language `#1'. Using the pattern for}%
\typeout{** the default language instead.}%
\else
\language=\csname l@#1\endcsname
\fi
#2}}
\providecommand{\BIBdecl}{\relax}
\BIBdecl

\bibitem{Springer1967}
M.~D. Springer and W.~E. Thompson, ``Bayesian confidence limits for the
  reliability of cascade exponential subsystems,'' \emph{IEEE Trans. Rel.},
  vol. R-16, no.~2, pp. 86--89, Sept. 1967.

\bibitem{Nadarajah2007}
S.~Nadarajah and S.~Kotz, ``Exact distribution of the peak runoff,'' \emph{J
  Hydrol (Amst)}, vol. 338, no.~3, pp. 325 -- 327, 2007.

\bibitem{Sumitomo2012}
Y.~Sumitomo and S.-H.~H. Tye, ``Stringy mechanism for a small cosmological
  constant,'' \emph{J. Cosmol. Astropart. Phys}, vol. 2012, no.~8, p. 032,
  2012.

\bibitem{Salo2006}
J.~Salo, H.~M. El-Sallabi, and P.~Vainikainen, ``{The distribution of the
  product of independent Rayleigh random variables},'' \emph{IEEE Trans.
  Antennas Propag.}, vol.~54, no.~2, pp. 639--643, Feb. 2006.

\bibitem{Karagiannidis2007}
G.~K. Karagiannidis, N.~C. Sagias, and P.~T. Mathiopoulos, ``{N*Nakagami}: A
  novel stochastic model for cascaded fading channels,'' \emph{IEEE Trans.
  Commun.}, vol.~55, no.~8, pp. 1453--1458, Aug. 2007.

\bibitem{Chizhik2002}
D.~Chizhik, G.~J. Foschini, M.~J. Gans, and R.~A. Valenzuela, ``{Keyholes,
  correlations, and capacities of multielement transmit and receive
  antennas},'' \emph{IEEE Trans. Wireless Commun.}, vol.~1, no.~2, pp.
  361--368, Apr. 2002.

\bibitem{Erceg1997}
V.~Erceg, S.~J. Fortune, J.~Ling, A.~J. Rustako, and R.~A. Valenzuela,
  ``{Comparisons of a computer-based propagation prediction tool with
  experimental data collected in urban microcellular environments},''
  \emph{IEEE J. Sel. Areas Commun.}, vol.~15, no.~4, pp. 677--684, May 1997.

\bibitem{Sofotasios2016}
S.~K. Yoo, S.~L. Cotton, P.~C. Sofotasios, and S.~Freear, ``Shadowed fading in
  indoor off-body communication channels: A statistical characterization using
  the $\kappa$-$\mu$/gamma composite fading model,'' \emph{IEEE Trans. Wireless
  Commun.}, vol.~15, no.~8, pp. 5231--5244, Aug. 2016.

\bibitem{Andrews2001}
M.~Al-Habash, L.~C. Andrews, and R.~L. Phillips, ``{Mathematical model for the
  irradiance probability density function of a laser beam propagating through
  turbulent media},'' \emph{Opt. Eng.}, vol.~40, no.~8, pp. 1554--1562, Aug.
  2001.

\bibitem{Zhong2015}
C.~Zhong, X.~Chen, Z.~Zhang, and G.~K. Karagiannidis, ``Wireless-powered
  communications: Performance analysis and optimization,'' \emph{IEEE Trans.
  Commun.}, vol.~63, no.~12, pp. 5178--5190, Dec. 2015.

\bibitem{Van2016}
P.-T. Van, H.-H.~N. Le, M.-D.~N. Le, and D.-B. Ha, ``Performance analysis in
  wireless power transfer system over {Nakagami} fading channels,'' in
  \emph{2016 Int. Conf. on Electron., Inform., and Commun. ({ICEIC} 2016).},
  Jan 2016.

\bibitem{Alcaraz2017}
O.~L. Alcaraz-L{\'o}pez, H.~Alves, R.~D. Souza, and E.~G. Fern{\'a}ndez,
  ``Ultrareliable short-packet communications with wireless energy transfer,''
  \emph{IEEE Signal Process. Lett.}, vol.~24, no.~4, pp. 387--391, Apr. 2017.

\bibitem{Kim2003}
D.~Kim, M.~A. Ingram, and W.~W. Smith, ``{Measurements of small-scale fading
  and path loss for long range RF tags},'' \emph{IEEE Trans. Antennas Propag.},
  vol.~51, no.~8, pp. 1740--1749, Aug. 2003.

\bibitem{Griffin2008}
J.~D. Griffin and G.~D. Durgin, ``Gains for {RF} tags using multiple
  antennas,'' \emph{IEEE Trans. Antennas Propag.}, vol.~56, no.~2, pp.
  563--570, Feb. 2008.

\bibitem{Odonoughue2012}
N.~O'Donoughue and J.~M.~F. Moura, ``On the product of independent complex
  gaussians,'' \emph{IEEE Trans. Signal Process.}, vol.~60, no.~3, pp.
  1050--1063, Mar. 2012.

\bibitem{Nakagami1960}
M.~Nakagami, ``The m-distribution- a general formula of intensity distribution
  of rapid fading,'' \emph{Stat. Methods Radio Propag.}, 1960.

\bibitem{AlouiniBook}
\BIBentryALTinterwordspacing
M.~K. Simon and M.-S. Alouini, \emph{{Digital Communication over Fading
  Channels}}.\hskip 1em plus 0.5em minus 0.4em\relax {Wiley-IEEE Press}, 2005.
  [Online]. Available: \url{http://www.worldcat.org/isbn/0471649538}
\BIBentrySTDinterwordspacing

\bibitem{Wang2003}
Z.~Wang and G.~B. Giannakis, ``A simple and general parameterization
  quantifying performance in fading channels,'' \emph{IEEE Trans. Commun.},
  vol.~51, no.~8, pp. 1389--1398, Aug. 2003.

\bibitem{Bhargav2018}
N.~Bhargav, C.~R.~N. da~Silva, Y.~J. Chun, {\'E}.~J. Leonardo, S.~L. Cotton,
  and M.~D. Yacoub, ``On the product of two $\kappa$-$\mu$ random variables and
  its application to double and composite fading channels,'' \emph{IEEE Trans.
  Wireless Commun.}, vol.~17, no.~4, pp. 2457--2470, April 2018.

\bibitem{Silva2018}
C.~R.~N. da~Silva, E.~J. Leonardo, and M.~D. Yacoub, ``Product of two envelopes
  taken from $\alpha$-$\mu$, $\kappa$-$\mu$, and $\eta$-$\mu$ distributions,''
  \emph{IEEE Trans. Commun.}, vol.~66, no.~3, pp. 1284--1295, March 2018.

\bibitem{Paris2014}
J.~F. Paris, ``Statistical characterization of $\kappa$-$\mu$ shadowed
  fading,'' \emph{IEEE Trans. Veh. Technol.}, vol.~63, no.~2, pp. 518--526,
  Feb. 2014.

\bibitem{Cotton2015}
S.~L. Cotton, ``Human body shadowing in cellular device-to-device
  communications: Channel modeling using the shadowed $\kappa$-$\mu$ fading
  model,'' \emph{IEEE J. Sel. Areas Commun.}, vol.~33, no.~1, pp. 111--119,
  Jan. 2015.

\bibitem{Laureano2015}
L.~Moreno-Pozas, F.~J. Lopez-Martinez, J.~F. Paris, and E.~Martos-Naya, ``The
  $\kappa$-$\mu$ shadowed fading model: Unifying the $\kappa$-$\mu $ and $\eta
  $-$\mu$ distributions,'' \emph{IEEE Trans. Veh. Technol.}, vol.~65, no.~12,
  pp. 9630--9641, Dec. 2016.

\bibitem{Lopez2017}
F.~J. Lopez-Martinez, J.~F. Paris, and J.~M. Romero-Jerez, ``The $\kappa$-$\mu$
  shadowed fading model with integer fading parameters,'' \emph{IEEE Trans.
  Veh. Technol.}, vol.~66, no.~9, pp. 7653--7662, Sept. 2017.

\bibitem{Leonardo2015}
E.~J. Leonardo and M.~D. Yacoub, ``The product of two $\alpha$-$\mu$ variates
  and the composite $\alpha$-$\mu$ multipath-shadowing model,'' \emph{IEEE
  Trans. Veh. Technol.}, vol.~64, no.~6, pp. 2720--2725, June 2015.

\bibitem{Bekkali2015}
A.~Bekkali, S.~Zou, A.~Kadri, M.~Crisp, and R.~V. Penty, ``Performance analysis
  of passive {UHF RFID} systems under cascaded fading channels and interference
  effects,'' \emph{IEEE Trans. Wireless Commun.}, vol.~14, no.~3, pp.
  1421--1433, Mar. 2015.

\bibitem{Gradstein2007}
I.~S. Gradshteyn and I.~M. Ryzhik, \emph{{Table of Integrals, Series and
  Products}}, 7th~ed.\hskip 1em plus 0.5em minus 0.4em\relax Academic Press
  Inc, 2007.

\bibitem{Erdelyi1954}
A.~Erd{\'e}lyi, W.~Magnus, F.~Oberhettinger, and F.~G. Tricomi, \emph{{Tables
  of Integral Transforms. {V}ol. {I}}}.\hskip 1em plus 0.5em minus 0.4em\relax
  McGraw-Hill Book Company, Inc., New York-Toronto-London, 1954.

\bibitem{Bi2016}
S.~Bi, Y.~Zeng, and R.~Zhang, ``{Wireless powered communication networks: an
  overview},'' \emph{IEEE Wireless Commun.}, vol.~23, no.~2, pp. 10--18, Apr.
  2016.

\bibitem{Romero2017}
J.~M. Romero-Jerez, F.~J. Lopez-Martinez, J.~F. Paris, and A.~J. Goldsmith,
  ``The fluctuating two-ray fading model: Statistical characterization and
  performance analysis,'' \emph{IEEE Trans. Wireless Commun.}, vol.~16, no.~7,
  pp. 4420--4432, July 2017.

\bibitem{Stockman1948}
H.~Stockman, ``Communication by means of reflected power,'' \emph{Proc. IRE},
  vol.~36, no.~10, pp. 1196--1204, Oct. 1948.

\bibitem{Griffin2010}
J.~D. Griffin and G.~D. Durgin, ``Multipath fading measurements at 5.8 {GHz}
  for backscatter tags with multiple antennas,'' \emph{IEEE Trans. Antennas
  Propag.}, vol.~58, no.~11, pp. 3693--3700, Nov. 2010.

\bibitem{Finkenzeller2010book}
K.~Finkenzeller, \emph{{RFID Handbook: Fundamentals and Applications in
  Contactless Smart Cards, Radio Frequency Identification and Near-Field
  Communication}}.\hskip 1em plus 0.5em minus 0.4em\relax John Wiley \& Sons,
  2010.

\bibitem{griffin2011fading}
J.~D. Griffin and G.~D. Durgin, ``{Fading Statistics for Multi-Antenna RF
  Tags},'' \emph{Handbook of Smart Antennas for RFID Systems}, p. 469, 2011.

\bibitem{abramowitz1964}
M.~Abramowitz and I.~A. Stegun, \emph{{Handbook of Mathematical Functions: With
  Formulas, Graphs, and Mathematical Tables}}.\hskip 1em plus 0.5em minus
  0.4em\relax Courier Corporation, 1964, vol.~55.

\end{thebibliography}

\flushend

\end{document}